\RequirePackage{lineno}
\documentclass[aps,pof,superscriptaddress,nofootinbib]{revtex4}
\usepackage[parfill]{parskip}    			
\usepackage{caption}					
\usepackage{subcaption}					
\usepackage{mathtools}
\usepackage{amssymb}
\usepackage{amsmath}
\usepackage{graphicx}					
\usepackage{epstopdf}					
\usepackage[latin1]{inputenc}				
\usepackage[T1]{fontenc}					

\usepackage{version}				
\usepackage{cases}						

\usepackage{pstricks}					

\usepackage{mathrsfs}
\usepackage[squaren, Gray]{SIunits}			
\usepackage{bm}
\usepackage[percent]{overpic}

\begin{document}

\title{A local approach to the study of energy transfers in incompressible magnetohydrodynamic turbulence}

\author{Denis Kuzzay}
\email{denis.kuzzay@obspm.fr}
\affiliation{
LESIA, Observatoire de Paris, Université PSL, CNRS, Sorbonne Université, Univ. Paris Diderot, Sorbonne Paris Cité, 5 place Jules Janssen, 92195 Meudon, France
}
\author{Olga Alexandrova}
\affiliation{
LESIA, Observatoire de Paris, Université PSL, CNRS, Sorbonne Université, Univ. Paris Diderot, Sorbonne Paris Cité, 5 place Jules Janssen, 92195 Meudon, France
}
\author{Lorenzo Matteini}
\affiliation{
LESIA, Observatoire de Paris, Université PSL, CNRS, Sorbonne Université, Univ. Paris Diderot, Sorbonne Paris Cité, 5 place Jules Janssen, 92195 Meudon, France
}

\begin{abstract}

We present a local approach to the study of scale-to-scale energy transfers in magnetohydrodynamic (MHD) turbulence. This approach is based on performing local averages of the physical fields, which amounts to filtering scales smaller than some parameter $\ell$. A key step in this work is the derivation of a local Kármán-Howarth-Monin relation which provides a local form of Politano and Pouquet's 4/3-law, without any assumption of homogeneity or isotropy. Our approach is exact, non-random, and we show its connection to the usual statistical results of turbulence. Its implementation on data obtained via a three dimensional direct numerical simulation of the forced, incompressible MHD equations from the John Hopkins turbulence database constitutes the main part of our study. First, we show that the local Kármán-Howarth-Monin relation holds well. The space statistics of local cross-scale transfers is studied next, their means and standard deviations being maximum at inertial scales, and their probability density functions (PDFs) displaying very wide tails. Events constituting the tails of the PDFs are shown to form structures of strong transfers, either positive or negative, and which can be observed over the whole available range of scales. As $\ell$ is decreased, these structures become more and more localized in space while contributing to an increasing fraction of the mean energy cascade rate. Finally, we highlight their quasi 1D (filament-like) or quasi 2D (sheet/ribbon-lke) nature, and show that they appear in areas of strong vorticity or electric current density.

\end{abstract}

\maketitle

\section{Introduction}


Among unsolved problems in classical physics, fully developed turbulence is a nonlinear and nonlocal phenomenon which has remained a great challenge \cite{RichardsonBook, Taylor1935-1, Taylor1935-2, KH1938, K41a, K41b, K41c, K41d, BatchelorBook, FrischBook, PopeBook}. Our understanding of turbulence is centered around the idea of an energy cascade, crucial to energy dissipation and the existence of a stationary flow at large scales \cite{RichardsonBook, K41d}. In the cascade picture, when a fluid is set in motion by some large-scale forcing, the input kinetic energy is transferred to small scales (direct cascade) due to nonlinear interactions, where it can be efficiently dissipated by viscous effects. Between the injection scale $L$ and the dissipative scales lies the inertial range where nonlinear forces govern the dynamics, and energy cascades without being directly injected or dissipated \cite{FrischBook}. Therefore, understanding the mechanisms which lead to this cascading process is of paramount importance to turbulence theory.

The idea of an energy cascade was formalized and linked to the laws of fluid dynamics, governed by the Navier-Stokes equations, around the 1940's. During this period, the mathematical tools developed by Taylor, von Kármán, Howarth, and Kolmogorov were rooted in a statistical view of turbulence, in which the fields at two points $\bm x$ and $\bm x'$ are described as correlated random variables. This approach culminated in the publication of four seminal papers \cite{K41a, K41b, K41c, K41d} which laid the foundations of Kolmogorov's 1941 theory of homogeneous, isotropic turbulence (referred to as K41 in the following). One crucial result was the derivation by von Kármán and Howarth of an equation governing the evolution of longitudinal autocorrelation of the velocity $\bm u$ \cite{KH1938}, which was later generalized by Monin without the isotropy assumption \cite{MoninYaglom, Monin1959}. The isotropic version of the Kármán-Howarth-Monin (KHM) relation served as a stepping stone to the formulation of K41 and the derivation of the famous 4/5-law

\begin{equation}
\label{4/5law}
\langle \delta_\ell u^3 \rangle = -\frac{4}{5}\epsilon_u\ell,
\end{equation}

where brackets denote ensemble averages and

\begin{equation}
\delta_\ell u\left(\bm x\right) \coloneqq \delta_\ell \bm u\left(\bm x\right) \cdot \frac{\bm \ell}{\ell} \coloneqq \left(\bm u\left(\bm x + \bm \ell\right) - \bm u\left(\bm x\right)\right) \cdot \frac{\bm \ell}{\ell},
\end{equation}


are the longitudinal velocity increments. In Eq. (\ref{4/5law}), the scale $\ell$ lies in the inertial range and $\epsilon_u$ is the mean kinetic energy dissipation rate. Kolmogorov's 4/5-law is a cornerstone of turbulence theory which has been thoroughly studied theoretically, experimentally, and numerically \cite{Lindborg1996, Sreenivasan1996, Antonia1997, Qian1997, Lindborg1999, Moisy1999, Danaila2002, Gotoh2002, Lundgren2002, Gagne2004, Antonia2006, Kaneda2008}. 
It formalizes the concept of energy cascade by connecting $\epsilon_u$, assumed to remain finite in K41, to the measurable third-order structure function. Moreover, it is the only scaling law which has been derived from Navier-Stokes equations, and is exact. Using the incompressibility condition, the 4/5-law may be expressed in terms of both longitudinal and vector increments as 

\begin{equation}
\label{4/3law}
\langle \delta_\ell u \vert \delta_\ell \bm u \vert^2 \rangle = -\frac{4}{3}\epsilon_u\ell,
\end{equation}


\textit{i.e.} a "4/3-law", which is the analog of the relation derived by Yaglom for temperature fluctuations in turbulent flows \cite{MoninYaglom, Yaglom1949}.



Interestingly enough, the 4/3-law may be generalized to magnetohydrodynamics (MHD) while keeping a simple expression such as in Eq. (\ref{4/3law}). In the presence of a magnetic field $\bm b$ (in Alfvén units), the 4/3-law is expressed for inertial scales in terms of the Elsässer variables $\bm z^\pm = \bm u \pm \bm b$ as

\begin{equation}
\label{4/3MHD1}
\langle \delta_\ell z^\mp \vert \delta_\ell \bm z^\pm\vert^2\rangle = -\frac{4}{3}\epsilon^\pm\ell,
\end{equation}

which formalizes the existence of a double direct cascade of $E^\pm \coloneqq \langle {\bm z^\pm}^2\rangle$ at rates $\epsilon^\pm$ or, equivalently, of both the total energy $E^T \coloneqq \langle \bm u^2 + \bm b^2 \rangle/2$ and the cross helicity $H^C \coloneqq \langle \bm u \cdot \bm b \rangle$ at rates $\epsilon^T = \left(\epsilon^+ + \epsilon^-\right)/2$ and $\epsilon^C = \left(\epsilon^+ - \epsilon^-\right)/2$ respectively. As in the hydrodynamic case, the derivation of relation (\ref{4/3MHD1}) is based upon the derivation of a generalization of the KHM relation to MHD assuming both homogeneity and isotropy \cite{PP1998b}. Note that it is possible to derive a law involving only the longitudinal components of the field \cite{PP1998b, PP1998a}. However, it takes a more complex form than (\ref{4/5law}), as it explicitly involves both third-order structure functions and third-order correlation functions of the longitudinal fields. Furthermore, the 4/5 factor is replaced by 8/15.

generalization of the 4/5-law can also be derived under the same assumptions. In order to do this, one may start from relation (\ref{4/3MHD1}) and use incompressibility, homogeneity and isotropy, or one may follow the steps of Kolmogorov as was first done in \cite{PP1998a}. However, the 4/5-law for MHD takes a more complex form than (\ref{4/5law}), as it explicitly involves both third-order structure functions and correlation functions of the longitudinal fields.

Recent theoretical advances in hydrodynamic turbulence have shown that the cascade phenomenology can be understood from a local, non-random description of energy transfers. Onsager, who contributed to turbulence theory during the 1940's \cite{Onsager1949, EyinkOnOnsager}, seems to have been the first to realize that the statistical notions presented above should hold for individual realizations of turbulent flows, \textit{i.e.} without any ensemble averages \cite{Eyink2003, Taylor2003}. These ideas were recently formalized in a clear mathematical framework by Duchon and Robert \cite{DR2000}, and a key step in describing the energy cascade from a local, non-statistical point of view is the derivation of a local version of the KHM relation. The local form of the 4/3-law follows after a few lines of calculations, where statistical averages are replaced by local averages over a ball of typical size $\ell$, and neither homogeneity nor isotropy are assumed (see also \cite{EyinkNotes, Eyink2008}). This approach, which allows for the generalization of the statistical picture presented above to a local description, was applied to empirical data in the framework of von Kármán flows \cite{Kuzzay2015}. Investigations of local scale-to-scale energy transfers led to the detection of extreme events at Kolmogorov scale, perhaps connected to singularities in the velocity field \cite{Saw2016, Kuzzay2017, MyThesis}. In the light of these positive results, further analyses were performed to study intermittency in the same context, and to examine atmospheric flows \cite{Debue2018,Faranda2018}. The developments brought about by this approach raise the question of what can be achieved in MHD turbulence, where local processes have been widely investigated in connection with turbulent energy dissipation \cite{Osman2011, Perri2012, Wu2013, Osman2014, Wan2016, Cerri2017, Franci2017, Perrone2017, Verdini2018}. Various approaches have been tested in order to gain some insight into the local physics of MHD flows, especially in the context of solar wind turbulence. However, these are either heuristic \cite{Sorriso2018a, Sorriso2018b}, or focus on second-order statistics rather than third-order \cite{Greco2009, Osman2012}. Recently, the approach introduced by Duchon and Robert \cite{DR2000} has been generalized to MHD \cite{Galtier2018}. This sets a theoretical background for the study of local energy transfers in turbulent magnetized flows, in connection with the dynamical equations.

The aim of this paper is to present the first investigations of this local approach from MHD data. In this study, we have chosen to work with numerical data from the John Hopkins turbulence database \cite{JHTDBlink, JHTDB1, JHTDB2, JHTDB3}. First, we present our own derivation of the local KHM relation for MHD flows in terms of Elsässer variables, largely following \cite{DR2000}, and discuss the physics of our approach. In this formulation, the connection to classical results of MHD turbulence \cite{PP1998a, PP1998b} is made and discussed. We then proceed to the study of the local KHM relation by checking that it holds in these data, and by checking that usual results of MHD turbulence are recovered globally over the whole flow. Next, we investigate the local organization of scale-to-scale energy transfers by providing maps at injection, inertial, and dissipative scales. The existence of structures of strong magnitude, either positive or negative, over the whole available range of scales is highlighted, and the statistics of cross-scale transfers is studied. In particular, we highlight that they get more and more localized as the scale $\ell$ is decreased, while contributing to an increasing fraction of the overall energy transfer rate. Moreover, we show that these structures are quasi 1D or 2D clusters of events constituting the tails of the probability density functions. Finally, we try to connect the existence of such structures with local strong gradients in the velocity and magnetic fields, and show a very good correlation with regions of strong vorticity or electric current density.

\section{A local approach}
\label{LocalApproach}

\subsection{The filtering approach}
\label{filtering_approach}


For our study of energy transfers at different scales, we use a local scale-by-scale approach which allows to separate different scales of motion. Let us start from the incompressible MHD equations for Elsässer variables $\bm z^\pm = \bm u \pm \bm b$

\begin{align}
\label{MHDeq}
\partial_t z^\pm_i + z^\mp_j \partial_j z^\pm_i &= -\partial_i p^\ast + \nu \partial_j\partial_j z^\pm_i + f_i,\\
\label{incomp}
\partial_j z^\pm_j &= 0,
\end{align}

where $p^\ast = p + \bm b^2/2$ is the sum of the hydrodynamic and magnetic pressures, $\bm f$ is some external forcing, and $\nu$ is the kinematic viscosity. Summation over repeated indices is implied. Since it does not change anything to our reasoning, we take the magnetic Prandtl number (ratio between the kinematic viscosity and magnetic resistivity) $P_m = 1$ for simplicity. The filtering approach is based on mollifying physical quantities with some kernel $G$ which is even, non-negative, spatially localized, has compact support on $\mathbb{R}^3$, and is such that $\int_{\mathbb{R}^3} d\bm r \ G\left(\bm r\right) = 1$. In order to formalize the notion of scale, we define the function $G_\ell$ such that $G_\ell\left(\bm r\right) \coloneqq \ell^{-3} G\left(\bm r/\ell\right)$ \cite{Germano1992}. Let us now define the coarse-grained Elsässer fields at scale $\ell$ by taking the convolution product of $\bm z^\pm$ with $G_\ell$

\begin{equation}
z^\pm_{i,\ell}\left(\bm x,t\right) \coloneqq \int_{\mathbb{R}^3} d\bm r \ G_\ell\left(\bm r\right) z^\pm_i\left(\bm x + \bm r,t\right).
\label{filtering}
\end{equation}

The filtered pressure field $p^\ast_\ell$ is defined in the same way. In Eq. (\ref{MHDeq}), the unknown Elsässer and pressure fields contain informations about the physics at all scales. The filtering process (\ref{filtering}) averages out fine details of the fields while keeping informations about scales larger than $\ell$. Therefore, $\bm z^\pm_\ell\left(\bm x,t\right)$ represents the average Alfvénic fluctuations over a volume of plasma of typical size $\ell$ at point $\bm x$ and time $t$. We thus see that this approach allows a local, scale-by-scale study of MHD turbulence, and will be well suited for our purposes.

\subsection{Local energy balance}
\label{localbalance}

We now want to derive a local energy balance at all scales based on the filtering approach presented in \ref{filtering_approach}. The following reasoning closely follows the work of \cite{DR2000}, where it was applied to hydrodynamics (see also \cite{EyinkNotes}). We define the energy of the Elsässer fields at scales larger than $\ell$ as 

\begin{equation}
\label{largescaleE}
e^\pm_\ell\left(\bm x,t\right) \coloneqq \frac{1}{2} z^\pm_i\left(\bm x,t\right) z^\pm_{i,\ell}\left(\bm x,t\right) = \int_{\mathbb{R}^3} d\bm r \ G_\ell\left(\bm r\right) \frac{1}{2} z^\pm_i\left(\bm x,t\right) z^\pm_i\left(\bm x + \bm r,t\right),
\end{equation}

where $z^\pm_i\left(\bm x,t\right) z^\pm_i\left(\bm x + \bm r,t\right)/2$ is called the point-split energy density of the $\bm z^\pm$ fields. When $\ell\to 0$, $e^\pm_\ell$ tends to the usual definition of the energy since $z^\pm_{i,\ell} \to z^\pm_i$. Note that $e^\pm_\ell$ can as well be interpreted as the local two-point autocorrelation of the Elsässer fields, averages being performed over a local volume of plasma of typical size $\ell$ at point $\bm x$ and time $t$. Let us also draw the attention of the reader to the fact that this definition for the large-scale energy is different from the one used when adopting a large-eddy simulation (LES) perspective $\mathcal{E}^\pm_\ell \coloneqq z^\pm_{i,\ell} z^\pm_{i,\ell}/2 = e^\pm_\ell + \left(\delta z^\pm_i\right)_\ell z^\pm_{i,\ell}/2$. In particular, $e^\pm_\ell$ is not guaranteed to be positive. However, it can be checked in data that it is mostly positive, especially at inertial and dissipative scales. The reasons for working with the alternate definition (\ref{largescaleE}) will be discussed in more details in Sec. \ref{linkKHM}. Let us now denote with a hat symbol the value of the fields taken at point $\bm x + \bm r$, \textit{i.e.} $\widehat{z^\pm_i} \coloneqq z^\pm_i\left(\bm x + \bm r,t\right)$ and $\widehat{p^\ast} \coloneqq p^\ast\left(\bm x + \bm r,t\right)$. Writing Eq. (\ref{MHDeq}) at point $\bm x$ multiplied by $\widehat{z^\pm_i}$ together with Eq. (\ref{MHDeq}) at point $\bm x + \bm r$ multiplied by $z^\pm_i$ and adding the two equations, it can be shown using the incompressibility condition (\ref{incomp}) that the point-split energy density $z^\pm_i\widehat{z^\pm_i}/2$ satisfies

\begin{multline}
\label{localKHMnofilter}
\partial_t \left(\frac{z^\pm_i\widehat{z^\pm_i}}{2} \right) \ + \ \partial_j \left[\frac{z^\pm_i\widehat{z^\pm_i}}{2} z^\mp_j  + \frac{p^\ast\widehat{z^\pm_j} + \widehat{p^\ast}z^\pm_j}{2} + \frac{\delta_r z^\mp_j \widehat{z^\pm_i}\widehat{z^\pm_i}}{4} - \nu\partial_j \left(\frac{z^\pm_i\widehat{z^\pm_i}}{2}\right) \right] \\
= \frac{1}{4} \partial_{r_j} \left(\delta_r z^\mp_j \delta_r z^\pm_i \delta_r z^\pm_i \right) - \nu\left(\partial_jz^\pm_i\right)\left(\partial_j\widehat{z^\pm_i}\right) + \frac{1}{2}\left(f_i \widehat{z^\pm_i} + \widehat{f}_i z^\pm_i \right),
\end{multline}

where $\partial_{r_j}$ denotes the derivative over the $j^{th}$ component of the displacement vector $\bm r$. Then, multiplying by $G_\ell(\bm r)$ and integrating over $\bm r$ yields the large-scale energy balance


\begin{equation}
\label{localKHM}
\partial_t e^\pm_\ell \ + \ \partial_j J^\pm_{j,\ell} = -\Pi^\pm_\ell - \mathcal{D}^\pm_{\nu,\ell} + \mathcal{F}^\pm_\ell,
\end{equation}

where

\begin{align}
\label{spaceflux}
J^\pm_{j,\ell} &\coloneqq e^\pm_\ell z^\mp_j  + \frac{1}{2}\left(p^\ast z^\pm_{j,\ell} + p^\ast_\ell z^\pm_j\right) + \frac{1}{4}\left(z^\mp_j z^\pm_i z^\pm_i \right)_\ell - \frac{1}{4}\left(z^\pm_i z^\pm_i \right)_\ell z^\mp_j - \nu\partial_j e^\pm_\ell,\\
\label{PiDR}
\Pi^\pm_\ell &\coloneqq \frac{1}{4} \int \ d\bm r \ \left[\partial_j G_\ell \left(\bm r\right)\right] \delta_r z^\mp_j \delta_r z^\pm_i \delta_r z^\pm_i,\\
\label{viscositylocal}
\mathcal{D}^\pm_{\nu,\ell} &\coloneqq \nu\left(\partial_jz^\pm_i\right)\left(\partial_jz^\pm_{i,\ell}\right),\\
\label{forcinglocal}
\mathcal{F}^\pm_\ell &\coloneqq \frac{1}{2}\left(f_i z^\pm_{i,\ell} + f_{i,\ell} z^\pm_i\right).
\end{align}

Eq. (\ref{localKHM}) is constituted, as usual, of a local time derivative of the large-scale energy $e^\pm_\ell$ and the divergence of a spatial energy flux $\bm J^\pm_\ell$ on the left-hand side, while the right-hand side gathers sink/source terms. $\Pi^\pm_\ell$ stems from nonlinear effects and thus describes the energy transfers through scale $\ell$. $\mathscr{D}^\pm_{\nu,\ell}$ describes viscous interactions, and $\mathscr{F}^\pm_\ell$ describes the energy injected by the external force $\bm f$ at scales larger than $\ell$. This result was first obtained in \cite{Galtier2018} (also following the derivation of \cite{DR2000}) where Eq. (\ref{localKHM}) was written with $\bm u$ and $\bm b$ instead of $\bm z^\pm$, and generalized to Hall-MHD\footnote{Actually, the first attempt at deriving Eq. (\ref{localKHM}) was made in \cite{Gao2013}. Even though it was published, this work contains many mistakes which lead to an incorrect result. The correct derivation was published in \cite{Galtier2018}}. However, the similarities with hydrodynamics appear more clearly in the Elsässer formulation, and taking $\bm b = 0$ directly yields the results obtained in \cite{DR2000}. In conclusion, since the large-scale energy balance (\ref{localKHM}) is local in both space and time, we will be able, from appropriate sets of data, to study the behaviour of each of these terms in various areas of MHD flows, at various scales. Finally, let us mention that local balances can also be derived for the large-scale kinetic and magnetic energy separately, using the point-split approach. Their derivation is provided in App. \ref{pourAluie}, but their detailed study is left for future works.

\subsection{Link to traditional results of turbulence}
\label{linkKHM}

Let us now discuss the connection of Eq. (\ref{localKHM}) to traditional, well-established results of turbulence. In homogeneous turbulence, it is possible, starting from Eq. (\ref{MHDeq}) and (\ref{incomp}), to derive an equation governing the evolution of the two-point autocorrelation function (or the mean point-split energy density) of the $\bm z^\pm$ variables

\begin{multline}
\label{KHMMHD}
\frac{1}{2}\partial_t \left<z^\pm_i\left(\bm x\right)z^\pm_i\left(\bm x+\bm \ell\right)\right> = \frac{1}{4} \partial_{\ell_i} \left<\delta z^\mp_i\left(\bm \ell\right)\vert\delta \bm z^\pm\left(\bm \ell\right)\vert^2\right> + \nu \partial_{\ell_j} \partial_{\ell_j} \left< z^\pm_i\left(\bm x\right)z^\pm_i\left(\bm x+\bm \ell\right)\right> \\
+ \frac{1}{2} \left< f_i\left(\bm x\right)z^\pm_i\left(\bm x+\bm \ell\right) + f_i\left(\bm x+\bm \ell\right) z^\pm_i\left(\bm x\right) \right>,
\end{multline}

where brackets denote ensemble averages. Eq. (\ref{KHMMHD}) was derived in \cite{PP1998b} assuming only homogeneity (without isotropy), and was written in an equivalent form in terms of increments. Its hydrodynamic counterpart ($\bm b = \bm 0$) is known as the Kármán-Howarth-Monin (KHM) relation, which is a name we will keep in the following to refer to Eq. (\ref{KHMMHD}). The derivation from Navier-Stokes equations also applies to MHD and can be found in \cite{FrischBook}, where it is shown to be an energy flux relation. The scale-to-scale transfer rate is identified with the first term on the right-hand side, stemming from nonlinear interactions, and expressed as a divergence over scales of third order structure functions. The 4/3-law of MHD

\begin{equation}
\label{4/3MHD}
\epsilon^\pm = - \frac{\langle \delta_\ell z^\mp \vert \delta_\ell \bm z^\pm\vert^2\rangle}{4/3\ell},
\end{equation}

follows from (\ref{KHMMHD}) in the inertial range by adding the assumptions of isotropy and stationarity according to K41 \cite{PP1998b}. $\epsilon^\pm$ is the mean dissipation rate of $E^\pm \coloneqq \langle {\bm z^\pm}^2\rangle$, assumed to be finite and positive in the limit of ideal MHD, and which is equal to the cascade rate by stationarity. More generally, we will define the quantity $\epsilon^\pm_\ell$ as

\begin{equation}
\epsilon^\pm_\ell \coloneqq - \frac{\langle \delta_\ell z^\mp \vert \delta_\ell \bm z^\pm\vert^2\rangle}{4/3\ell},
\end{equation}

at any scale $\ell$ from injection to dissipative scales. In particular, $\epsilon^\pm_\ell = \epsilon^\pm$ for $\ell$ lying in the inertial range.

%
%
%
%

%
%
%
%
%

Going back to the results derived in Sec. \ref{localbalance}, we see that Eq. (\ref{localKHM}) may be interpreted as a local version of the KHM relation for MHD (\ref{KHMMHD}), where ensemble averages are replaced by local averages over a local volume of typical size $\ell$. The presence of the divergence term on the left-hand side of (\ref{localKHM}) comes from the local inhomogeneity of the spatial flux, and vanishes after averaging using the statistical homogeneity assumption. Note that our local approach allowed us to derive Eq. (\ref{localKHM}) without any assumption on homogeneity or isotropy. The correspondence between terms in Eq. (\ref{localKHM}) and (\ref{KHMMHD}) is easily made, and the KHM relation is recovered from its local version by averaging over a statistical ensemble, assuming homogeneity. Therefore, Eq. (\ref{localKHM}) is a stronger result than (\ref{KHMMHD}). 

If we now focus on the expression of $\Pi^\pm_\ell$ given in (\ref{PiDR}), we see that local energy transfers through scales are still related to the divergence of the cube of increments. In (\ref{PiDR}), an integration by part has been made in order to make the gradient operator act on the test function $G_\ell$ (boundary terms vanish since $G$ has compact support) rather than directly on the increments. However, the most interesting result is perhaps to note that the 4/3-law (\ref{4/3MHD}) has a local counterpart. Indeed, generalizing the derivation in \cite{DR2000} (see also \cite{EyinkNotes}) to MHD, it can be shown after a few lines of calculation that in the case of a spherically symmetric filter kernel ($G\left(\bm r\right) = G\left(r\right)$), the expression of $\Pi^\pm_\ell$ may be simplified into

\begin{equation}
\label{4/3local}
\Pi^\pm_\ell = - \frac{\langle \delta_\ell z^\mp \vert \delta_\ell \bm z^\pm\vert^2\rangle_{ang}}{4/3\ell}, 
\end{equation}

for $\ell$ lying in the inertial range. The quantity

\begin{equation}
\langle \delta_\ell z^\mp \vert \delta_\ell \bm z^\pm\vert^2\rangle_{ang} \coloneqq \frac{1}{4\pi} \int_0^\pi \sin\theta d\theta \int_0^{2\pi} d\varphi \ \delta_\ell z^\mp \vert \delta \bm z^\pm\vert^2,
\end{equation}

is the angular average of $\delta_\ell z^\mp \vert \delta_\ell \bm z^\pm\vert^2$ over the unit sphere in three dimensions. This result was first given in \cite{DR2000} for the hydrodynamic case. Of course, $\Pi^\pm_\ell$ should be independent of $\ell$ in a well-defined inertial range. Let us stress that the derivation of (\ref{4/3local}) presented here is very similar to the derivation of (\ref{4/3MHD}) in \cite{PP1998b}, in that both are based on a KHM relation in order to arrive at an exact scaling relation for third order structure functions. The main difference between the two approaches lies in the type of averages that are performed. It is very interesting to note that the 4/3-law, which formalizes the concept of a dual cascade in MHD, still holds in a local (in space and time) sense. As stated in \cite{EyinkNotes}, the results presented here are much stronger than the usual statistical ones, because they hold for individual realizations of MHD flows. Indeed, the reasoning leading to Eq. (\ref{localKHMnofilter}) - (\ref{forcinglocal}) is exact, and Eq. (\ref{4/3local}) follows for any individual solution to the incompressible MHD equations, without performing any ensemble averages, and without making any assumptions on homogeneity or isotropy. For more discussions, see \cite{EyinkNotes, Onsager1949, Eyink2008, EyinkOnOnsager} which focus on hydrodynamics, but can be generalized to MHD.

Finally, let us highlight the fact that space filtering has attracted a lot of interest in the past few years, in order to investigate the local physics of turbulent MHD flows. For instance, separate energy balances for the kinetic and magnetic energies have been derived in order to study the scale locality of the double cascade, the effect of subscale terms, and kinetic scales \cite{Aluie2010, Aluie2017, Yang2017, Bian2019}. This approach has also been used in Hall-MHD numerical data in order to quantify the correlation between regions of nonzero scale-to-scale energy transfers and the presence of coherent structures in these areas \cite{Camporeale2018}. In these studies, the authors adopt a LES approach in which the large-scale energy is defined as the square of the filtered fields $\mathcal{E}^\pm_\ell \coloneqq z^\pm_{i,\ell} z^\pm_{i,\ell}/2 = e^\pm_\ell + \left(\delta z^\pm_i\right)_\ell z^\pm_{i,\ell}/2$ (to be compared with Eq. (\ref{largescaleE})). The main difference between the two approaches comes from two alternate definitions of the large-scale energy which lead to two different definitions for the nonlinear energy flux \cite{EyinkNotes}. Indeed, from a LES approach, the local scale-to-scale transfers are described by $\tau^\pm_{ij,\ell}\partial_j z^\pm_{i,\ell}$, where $\tau^\pm_{ij,\ell} = \left(z^\pm_i z^\mp_j \right)_\ell - z^\pm_{i,\ell}z^\mp_{j,\ell}$. As noted by Onsager (see [35]), this flux depends on increments only since

\begin{equation}
\tau^\pm_{ij,\ell} = \int \ d\bm r \ G_\ell \left(\bm r\right) \delta_r z^\pm_i \delta_r z^\mp_j - \int \ d\bm r \ G_\ell \left(\bm r\right) \delta_r z^\pm_i \int \ d\bm r \ G_\ell \left(\bm r\right) \delta_r z^\mp_j,
\end{equation}

and

\begin{equation}
\partial_j z^\pm_{i,\ell} = -\int \ d\bm r \ \left[\partial_j G_\ell \left(\bm r\right)\right] \delta_r z^\pm_i.
\end{equation}

As a consequence, it can be seen that the local scale-to-scale energy flux obtained from the LES perspective is not straightforwardly related to $\Pi^\pm_\ell$ defined in Eq. (\ref{PiDR}) because local averages on increments are taken in a different way. Note, however, that these can be considered as a different way of defining local structure functions. Therefore, the choice of definition for the large-scale energy depends on which type of structure functions one chooses to study, some of them being more directly related to energy flux (as is the case in our study, see \cite{EyinkNotes, Eyink2008} for a more detailed discussion). 

\section{Application to turbulent numerical data}


Direct numerical simulations (DNS) are a powerful tool to investigate MHD flows. They provide a framework in which the main features of the flow such as the forcing, the kinematic and magnetic viscosities, or the initial and boundary conditions may be precisely controlled. As such, they have been extensively used for the study of MHD turbulence, under a wide range of conditions \cite{Zikanov1998, Mininni2006, Dmitruk2009, Mininni2009, Dallas2014, Linkman2015, Hellinger2018}. In the end, DNS give access to all components of the relevant fields on a grid. Spatial gradients as well as filtered quantities can therefore be readily computed at several scales, which will enable us to visualize local energy transfers in various areas, at various $\ell$, at the same time. As a consequence, DNS allow for investigations which are not possible from spacecraft data, organized as time series. For such data, we may interpret time scales as space scales using Taylor's hypothesis, but only one-dimensional informations are recovered about the flow. Even multi-spacecraft missions such as CLUSTER or MMS \cite{cluster, mms}, which aim at investigating the local three-dimensional structure of the solar wind and the Earth's magnetoshpere, are limited to one region of space at a time, and a single scale (typically the distance between the spacecrafts). DNS provide a tool to investigate the three-dimensional structure of MHD flows on many scales without using Taylor's hypothesis. Consequently, they are well suited to perform our study and encourage future applications to plasma physics.

\subsection{Simulation parameters and data characteristics}
\label{DNSparameters}

\begin{figure}	
	\centering
	\begin{subfigure}[t]{8.5cm}
		\caption{}
		\label{}
		\begin{overpic}[width=8.5cm]{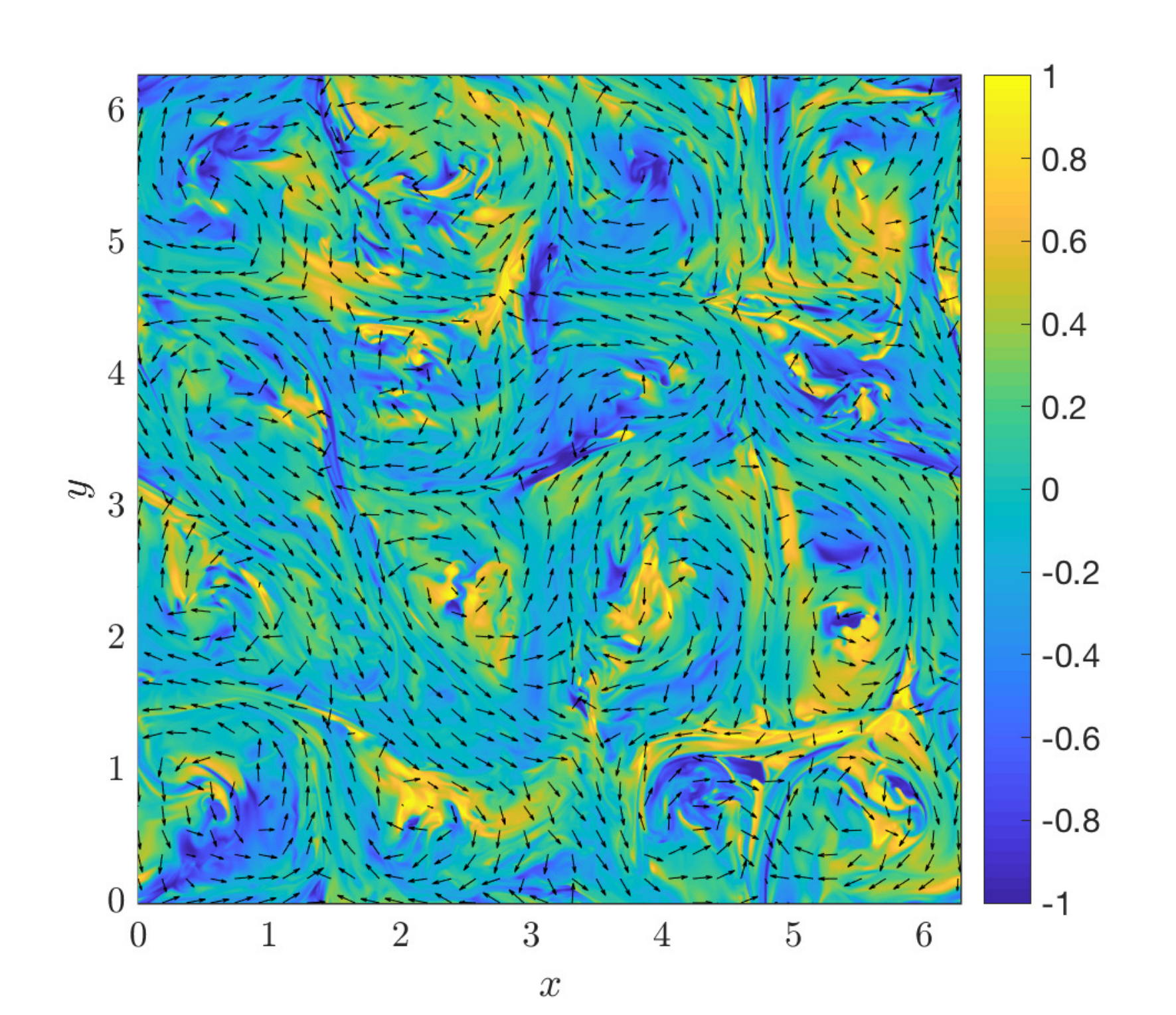}
       \put(70,83){\large $\bm u$}
       \end{overpic}
	\end{subfigure}
	\quad
	\begin{subfigure}[t]{8.5cm}
		\caption{}
		\label{}
		\begin{overpic}[width=8.5cm]{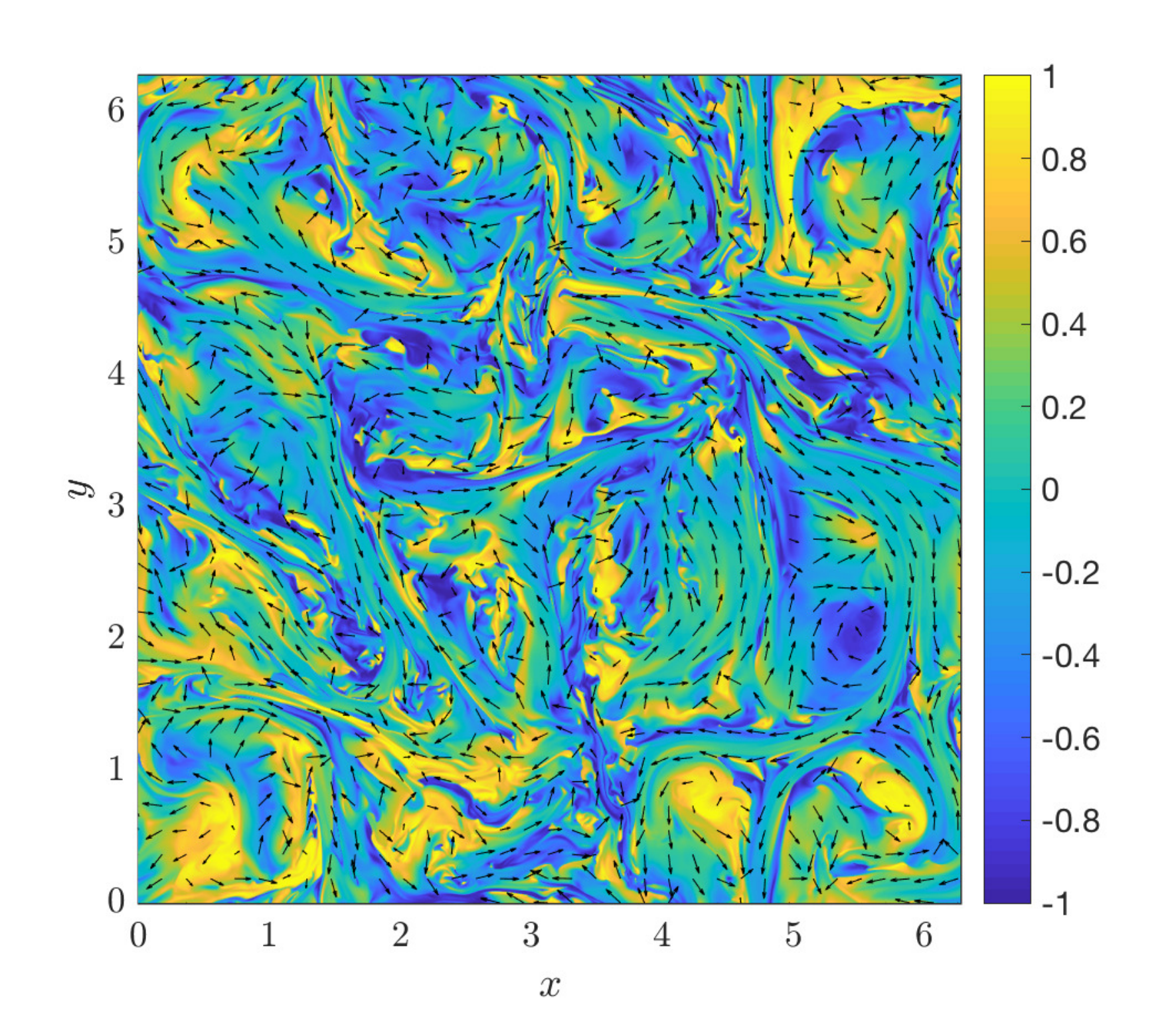}
       \put(20,83){\large $\bm b$}
       \end{overpic}
	\end{subfigure}
	\begin{subfigure}[t]{8.5cm}
		\caption{}
		\label{}
		\begin{overpic}[width=8.5cm]{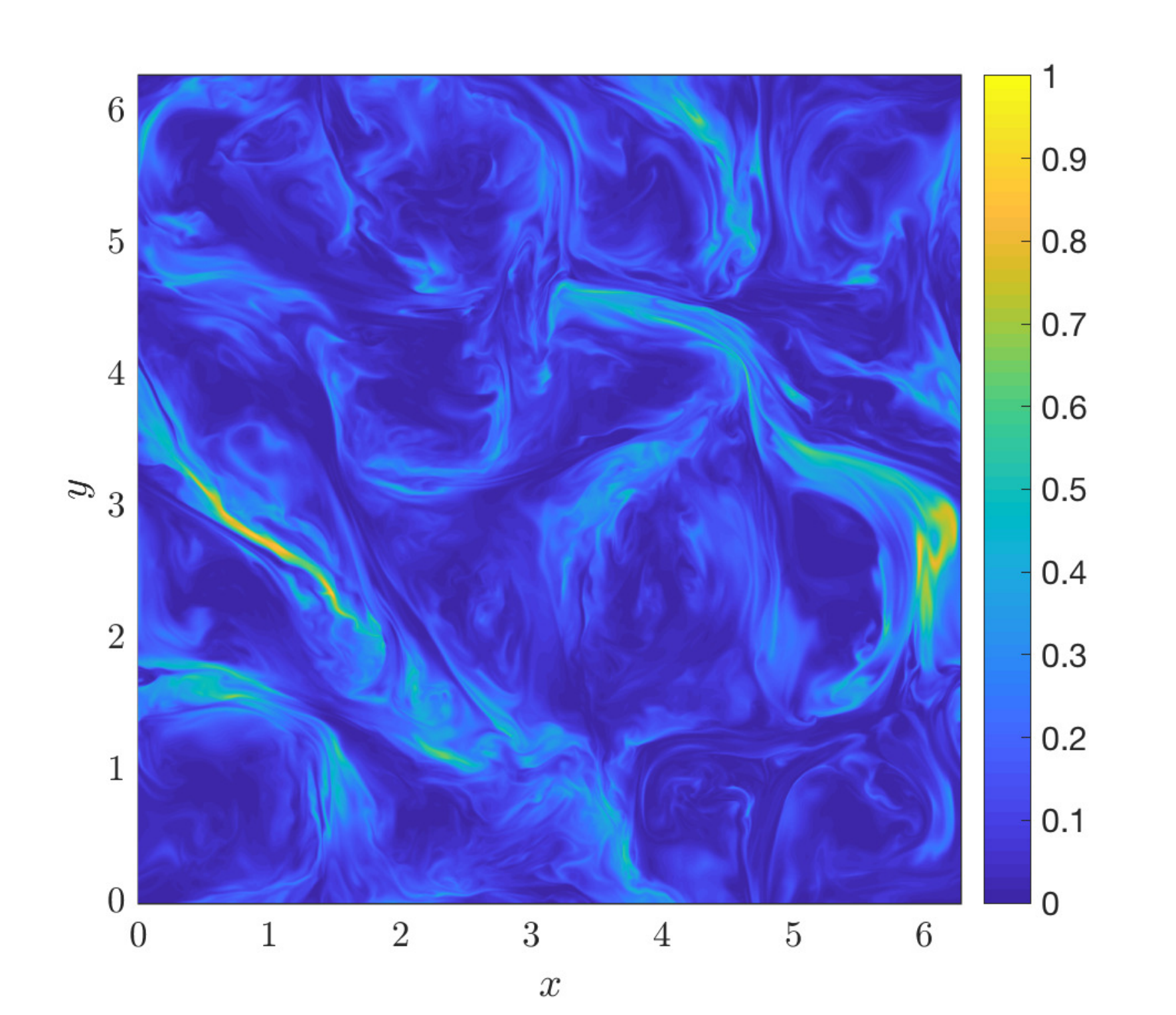}
       \put(70,83){\large $E_u$}
       \end{overpic}
	\end{subfigure}
	\quad
	\begin{subfigure}[t]{8.5cm}
		\caption{}
		\label{}
		\begin{overpic}[width=8.5cm]{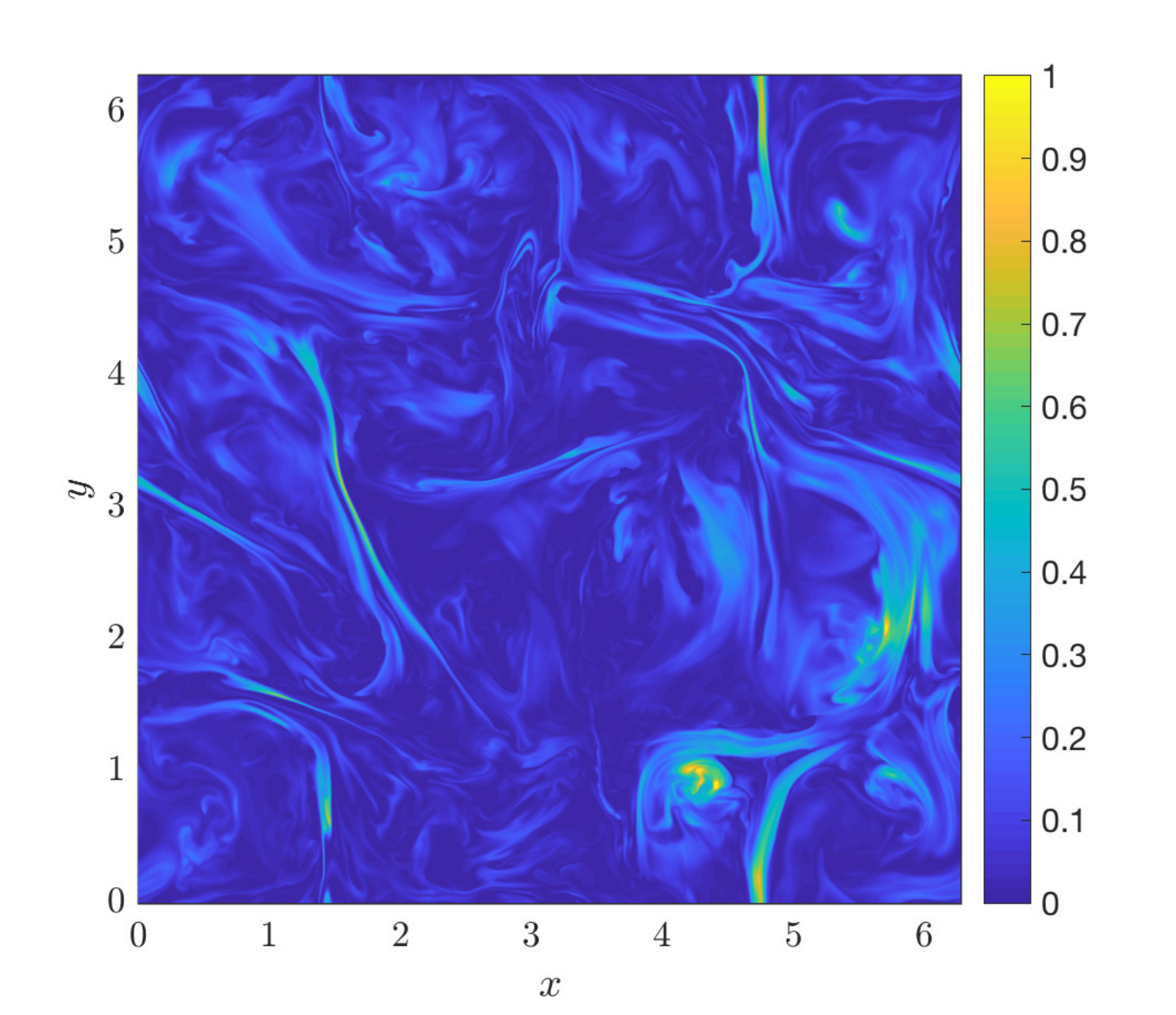}
       \put(20,83){\large $E_b$}
       \end{overpic}
	\end{subfigure}
		\begin{subfigure}[t]{8.5cm}
		\caption{}
		\label{}
		\begin{overpic}[width=8.5cm]{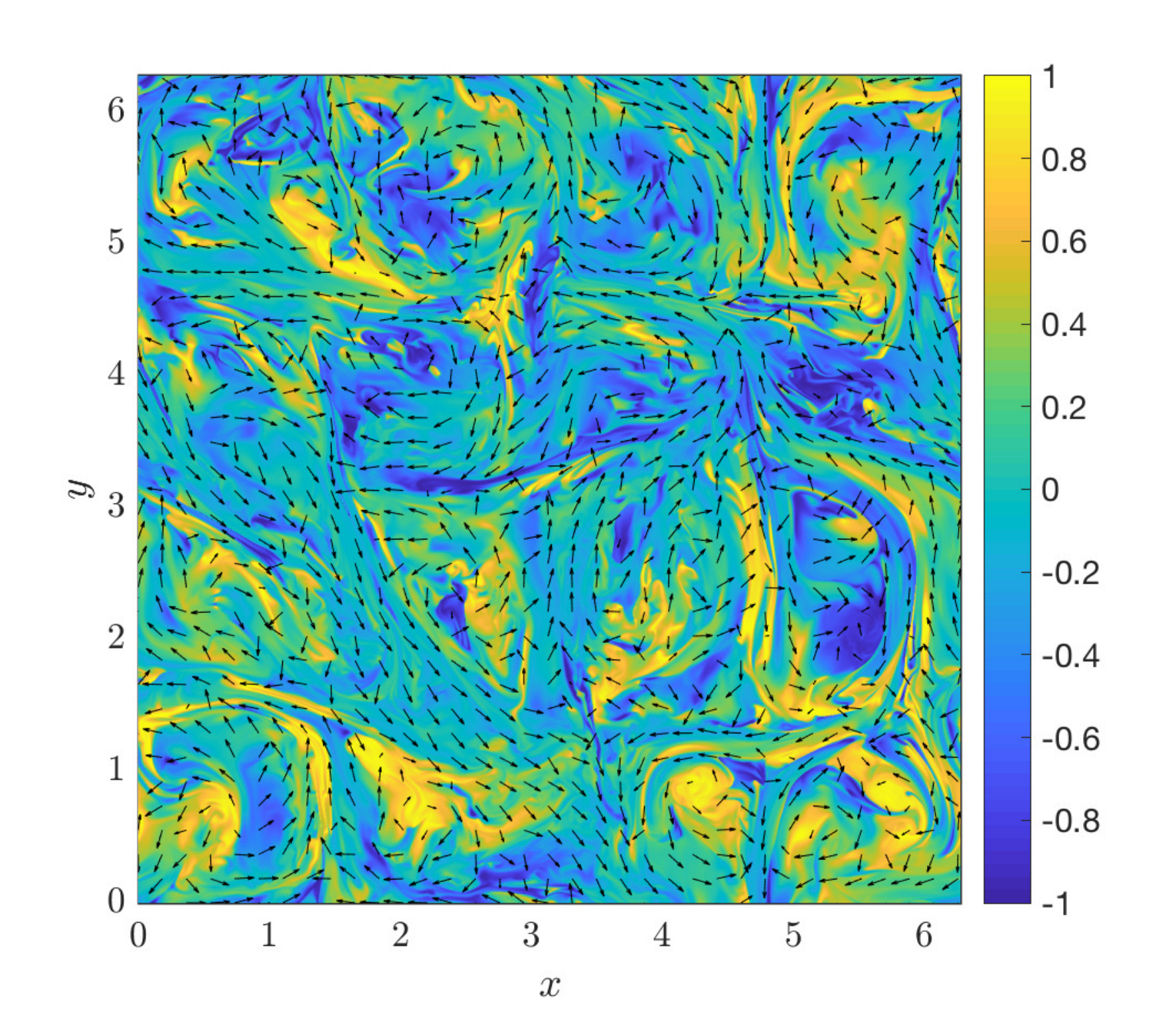}
       \put(70,83){\large $\bm z^+$}
       \end{overpic}
	\end{subfigure}
	\quad
	\begin{subfigure}[t]{8.5cm}
		\caption{}
		\label{}
		\begin{overpic}[width=8.5cm]{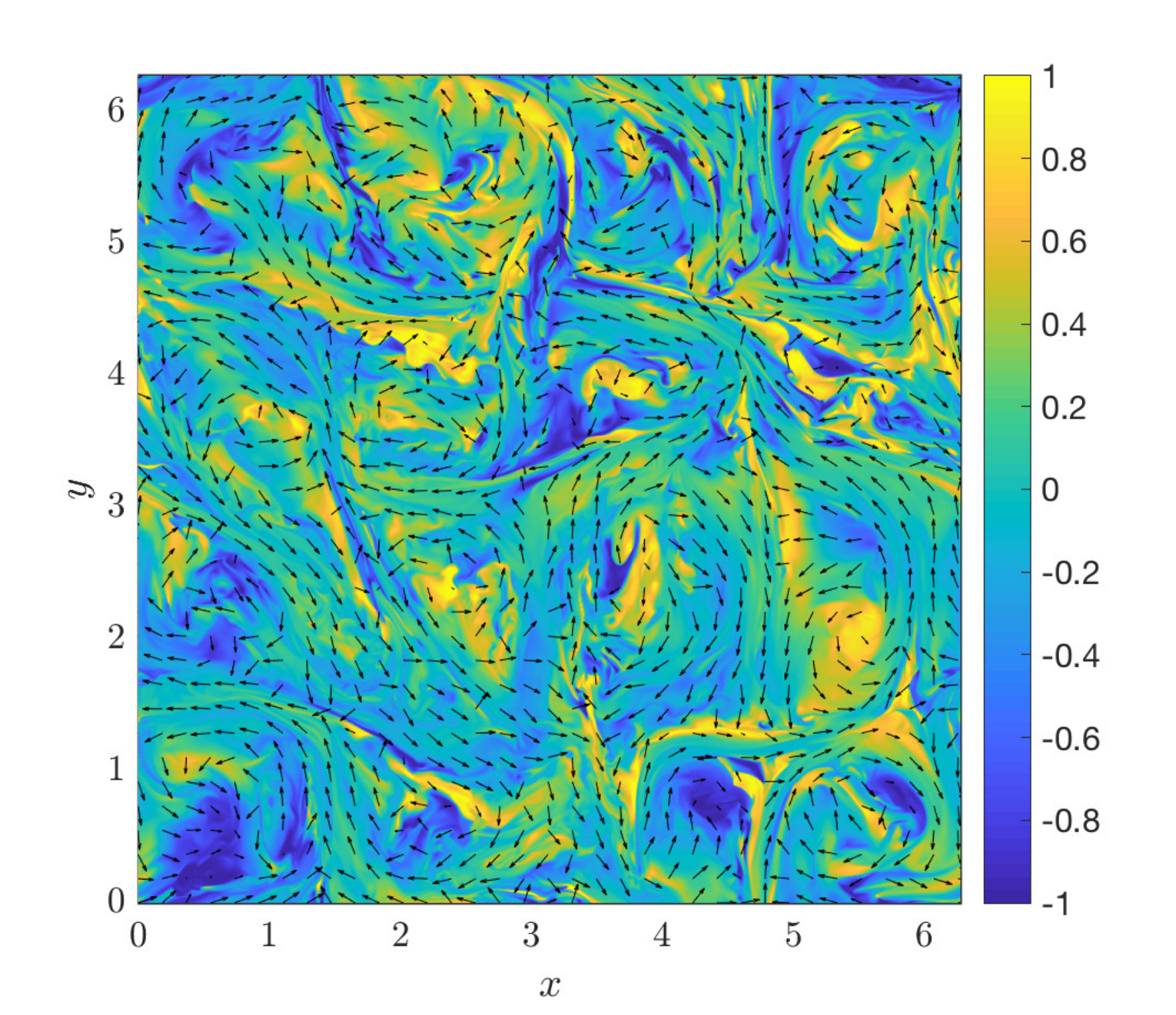}
       \put(20,83){\large $\bm z^-$}
       \end{overpic}
	\end{subfigure}
	\caption{Typical slices of the physical fields in the $\left(xy\right)$ plane of forcing for the snapshot considered in our analysis. For vector fields: arrows represent the in-plane component and the color map represents the out-of-plane component. The vector fields have been normalized by their norm.}
	\label{Fields}
\end{figure}

In the following, we use forced MHD turbulence data downloaded from the open John Hopkins Turbulence Database \cite{JHTDBlink}. The data were generated by a DNS of the 3D incompressible MHD equations, in a cubic domain of size $2\pi$ with periodic boundary conditions, and resolution $1024^3$. As in Sec. \ref{LocalApproach}, the magnetic Prandtl number is unity with $\nu = 1.1 \times 10^{-4}$. The flow is forced at large scales in the $\left(xy\right)$ plane by a steady Taylor-Green body force added to the momentum (Navier-Stokes) equation

\begin{equation}
\bm f = f_0 \left[ \sin\left(k_fx\right) \cos\left(k_fy\right) \cos\left(k_fz\right) \bm{e_x} - \cos\left(k_fx\right) \sin\left(k_fy\right) \cos\left(k_fz\right) \bm{e_y} \right],
\label{forcing}
\end{equation}

where $f_0 = 0.25$ and $k_f = 2$, and we define $L = \pi/2$ as the energy injection scale. Data have been stored after the flow had reached a statistically stationary regime, and may be accessed remotely through a web-service interface \cite{JHTDB1, JHTDB2, JHTDB3}. For our analysis, we have downloaded one snapshot, typical slices of $\bm u$, $\bm b$, $E_u \coloneqq \bm u^2/2$, $E_b \coloneqq \bm b^2/2$, and $\bm z^\pm$ in the $\left(xy\right)$ plane being displayed in Fig. \ref{Fields}. The vector fields exhibit characteristic features of turbulent flows, appearing as disordered with vortical structures, and the spatial distributions of the kinetic and magnetic energy densities are inhomogeneous, as expected. Moreover, the spatial distributions of $\bm z^+$ and $\bm z^-$ are very patchy, consistent with that of $\bm u$ and $\bm b$, and we will see that they lead to an equally inhomogeneous contribution of the local nonlinear couplings governing the cascade. In the stationary regime, the mean kinetic and magnetic energy dissipation rates respectively are $\epsilon_u = 1.1 \times 10^{-2}$ and $\epsilon_b = 2\epsilon_u$. The injection, cascade, and dissipation rates of total energy are all equal to $\epsilon^T = \epsilon_u + \epsilon_b = 3.3 \times 10^{-2}$. The corresponding Taylor-scale Reynolds numbers\footnote{We recall that the Taylor microscale is defined as $\lambda_w = (15\nu/\epsilon_w)^{1/2}w'$, where $w$ denotes either $u$ or $b$, and $w'$ is the root mean square of the fluctuations of $w$.} fluctuate around $Re_\lambda^u = 186$ and $Re_\lambda^b = 144$. The ratio of magnetic to kinetic energy is $\chi_M = \langle b^2\rangle/\langle u^2\rangle = 1.1$, and the $\left(\bm u, \bm b \right)$ correlation is weak $\langle\bm u\cdot\bm b\rangle/\langle \bm u^2 + \bm b^2\rangle \approx 10^{-2}$. The energy spectra cover around two decades of scales, with one full decade where they closely follow power laws \cite{JHTDBlink}. Note that from the DNS parameters, we can already see that the statistical properties of both Elsässer variables will be almost identical. Indeed, the Prandtl number being unity, the forcing being the same for $\bm z^+$and $\bm z^-$, and the injection rate of cross helicity being small, it can be anticipated that both variables will have a statistically symmetric role. This can be noticed on the energy spectra \cite{JHTDBlink}, and will be confirmed during our study. More details about the DNS may be found at \cite{JHTDBlink}.

\subsection{Data processing method}

In the following study, all gradients and convolution products are computed by making use of the periodic boundary conditions through fast Fourier transforms. The filtering kernel $G$ is defined as follows

\begin{equation}
  G\left(r\right)=
     \begin{cases}
        \frac{1}{\mathcal{N}} \exp\left(\frac{-1}{1-r^2}\right) & \text{if $r < 1$}, \\
        0  & \text{elsewhere},
     \end{cases}
\end{equation}

where $\mathcal{N}$ is a normalizing constant ensuring that $\int_{\mathcal{V}} d\bm r \ G\left(\bm r\right) = 1$, $\mathcal{V}$ being the volume of the whole cube. The function $G$ satisfies all the properties given in Sec. \ref{filtering_approach}, and is spherically symmetric. Let us note that for any finite scale $\ell$, filtered quantities depend on the choice of $G$, while their limits for $\ell \to 0$ do not \cite{Galtier2018}. However, physical results are expected to be independent (or, at least, weakly dependent) on the choice of kernel as long as it has the same symmetries as $G$ defined here, and the properties given in Sec. \ref{filtering_approach} are satisfied.

Finally, note that $\Pi^\pm_\ell$ as written in (\ref{PiDR}) is not a convolution product. However, by developing the increments under the integral it can be expressed as the sum of four terms where convolution products appear \cite{DR2000}

\begin{equation}
\Pi^\pm_\ell = - \partial_i\left(z^\mp_i z^\pm_j z^\pm_j\right)_\ell + 2 z^\pm_j \partial_i\left(z^\mp_i z^\pm_j \right)_\ell - 2 z^\mp_i z^\pm_j \partial_i\left( z^\pm_j\right)_\ell + z^\mp_i \partial_i\left( z^\pm_j z^\pm_j\right)_\ell.
\end{equation}

\subsection{Check local and global balance}
\label{CheckBalance}


\begin{figure}	
	\centering
	\begin{subfigure}[t]{8.5cm}
		\caption{}
		\label{BalanceFigurea}
		\includegraphics[width=9.1cm]{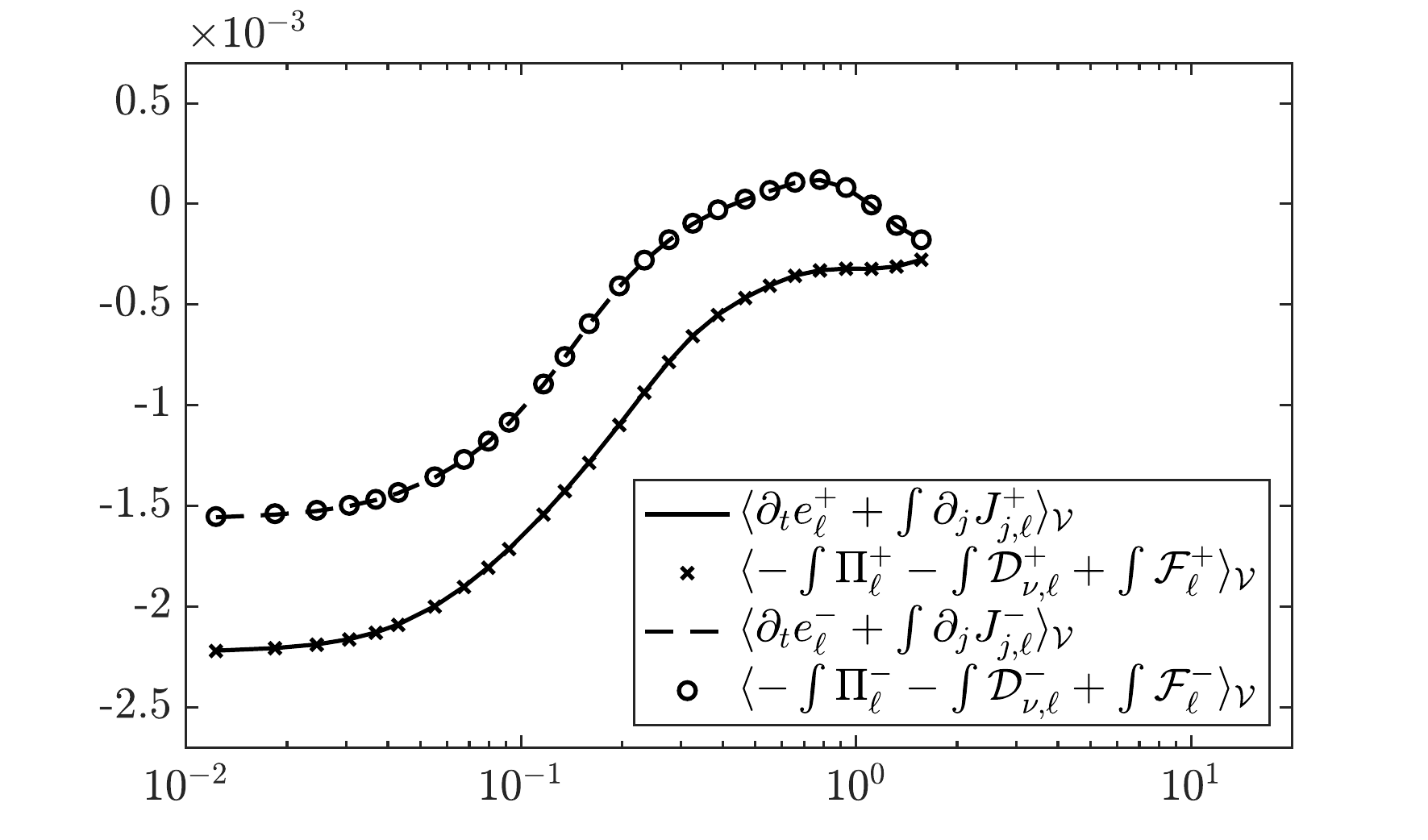}			
	\end{subfigure}
	\quad
	\begin{subfigure}[t]{8.5cm}
		\caption{}
		\label{BalanceFigureb}
		\includegraphics[width=9.1cm]{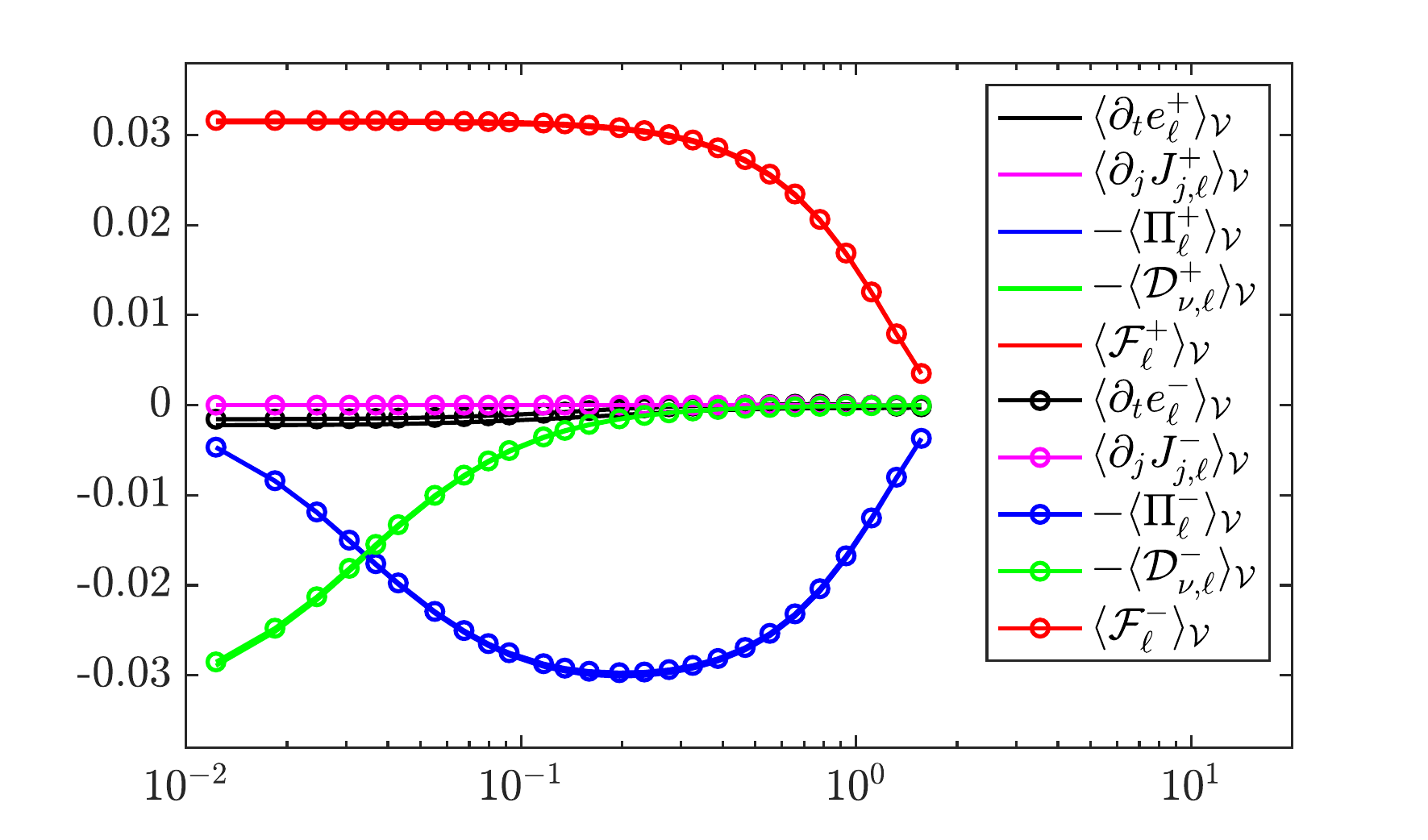}	
	\end{subfigure}
		\begin{subfigure}[t]{8.5cm}
		\caption{}
		\label{BalanceFigurec}
		\includegraphics[width=9.1cm]{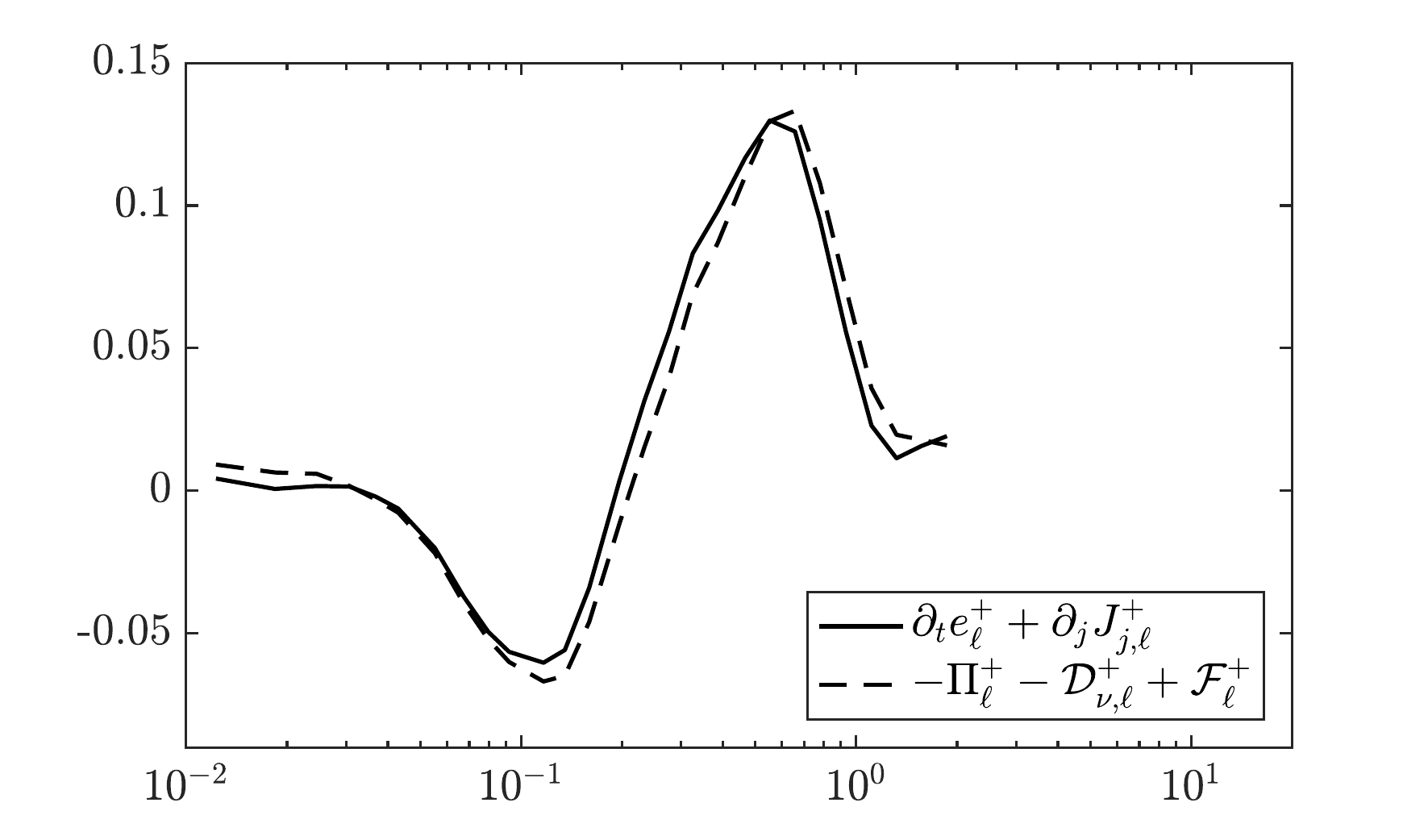}			
	\end{subfigure}
	\quad
	\begin{subfigure}[t]{8.5cm}
		\caption{}
		\label{BalanceFigured}
		\includegraphics[width=9.1cm]{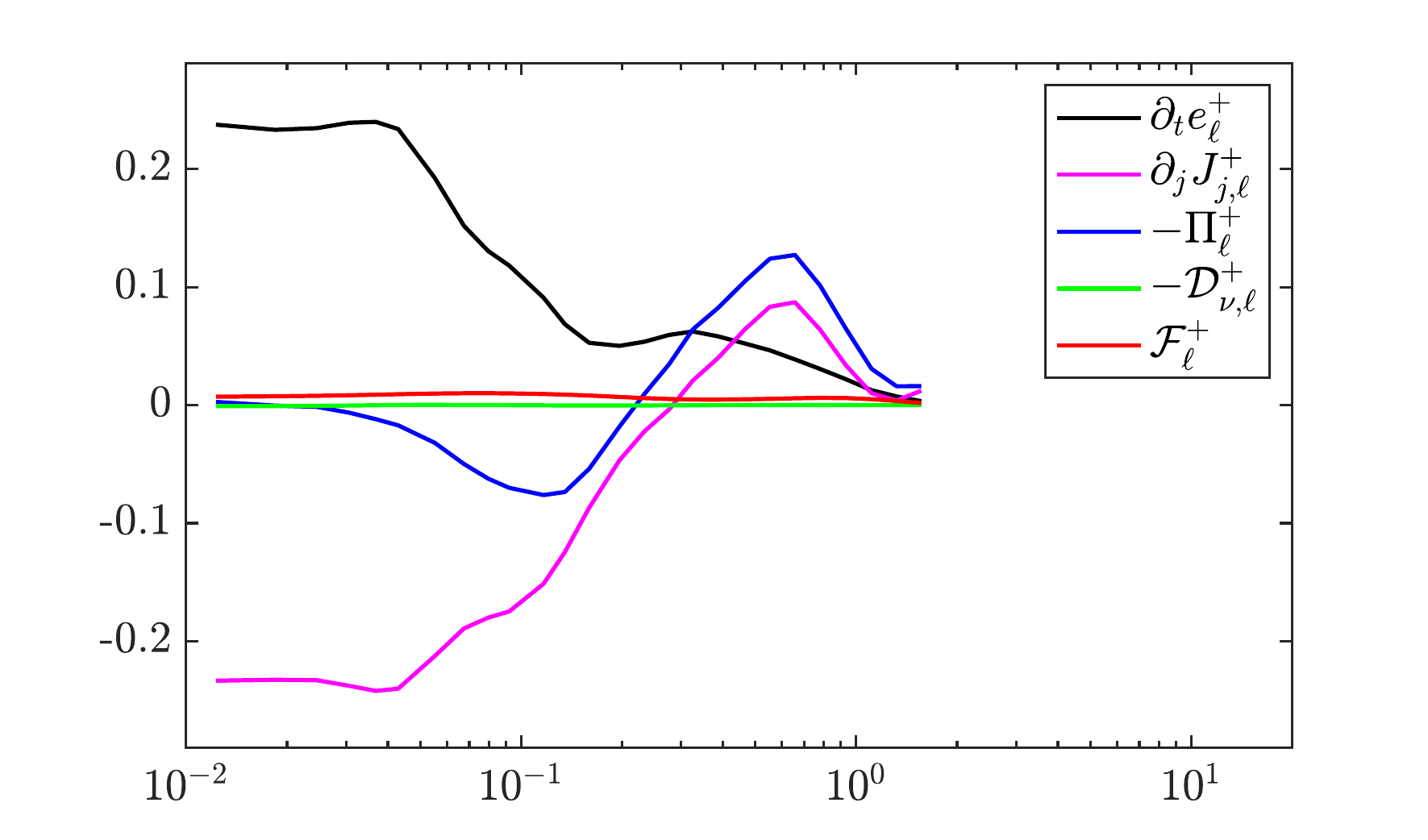}	
	\end{subfigure}
		\begin{subfigure}[t]{8.5cm}
		\caption{}
		\label{BalanceFiguree}
		\includegraphics[width=9.1cm]{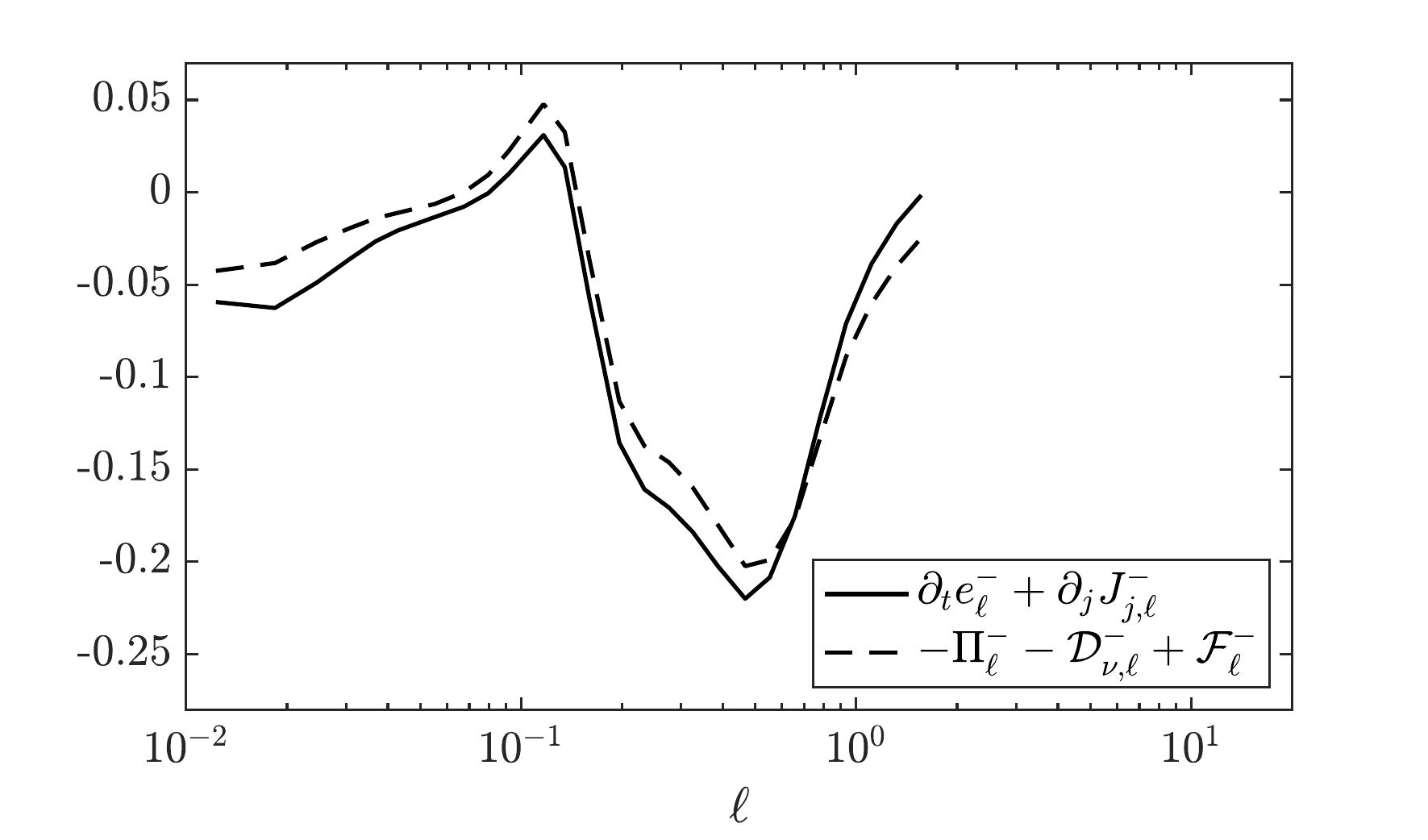}			
	\end{subfigure}
	\quad
	\begin{subfigure}[t]{8.5cm}
		\caption{}
		\label{BalanceFiguref}
		\includegraphics[width=9.1cm]{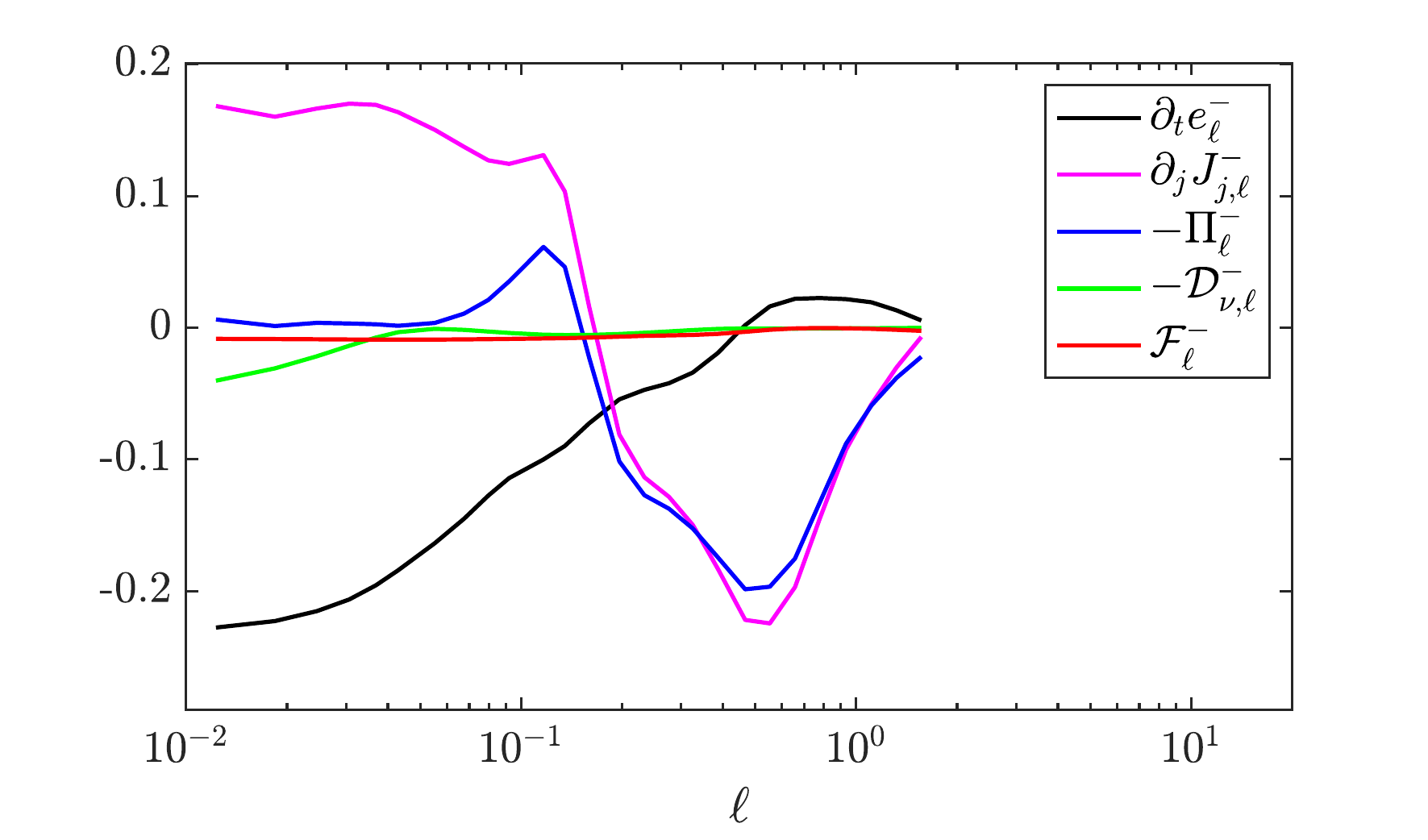}	
	\end{subfigure}
	\caption{Kármán-Howarth-Monin relation as a function of scale $\ell$. (a) balance between the left and right hand sides of Eq. (\ref{KHMglobal}). (b) contribution of each terms constituting (\ref{KHMglobal}) to the balance displayed in (a). (c) and (e) local balance between the left and right hand sides of Eq. (\ref{localKHM}) at one same point. (d) and (f) contribution of each terms constituting (\ref{localKHM}) to the balance displayed in (c) and (e) respectively.}
	\label{BalanceFigure}
\end{figure}

In Sec. \ref{LocalApproach}, we have shown that a key result for the study of local energy transfers was the derivation of two local KHM relations for $\bm z^+$ and $\bm z^-$ (see Eq. (\ref{localKHM}) - (\ref{forcinglocal})). Here, we want to check whether these relations hold globally over the whole box, as well as locally at grid points. This is what is shown on Fig. \ref{BalanceFigure}. Globally, the space-averaged KHM relations read

\begin{equation}
\label{KHMglobal}
\langle\partial_t e^\pm_\ell\rangle_\mathcal{V} = -\langle\Pi^\pm_\ell\rangle_\mathcal{V} - \langle\mathcal{D}^\pm_{\nu,\ell}\rangle_\mathcal{V} + \langle\mathcal{F}^\pm_\ell\rangle_\mathcal{V},
\end{equation}

where $\langle\cdot\rangle_\mathcal{V}$ denotes space averaging and the divergence term vanishes due to the periodic boundary conditions. Fig. \ref{BalanceFigurea} displays the variations of the left and right-hand sides (respectively denoted as LHS and RHS in the following) of Eq. (\ref{KHMglobal}) as a function of scales. As can be seen, the curves representing the LHS and RHS of both equations are undistinguishable for both balances. This means that the KHM relations hold very well globally. This is not surprising since Eq. (\ref{KHMglobal}) expresses that for every scale $\ell$, a time variation of the total energy of the Elsässer fields can only come from a difference between the total injected power and the joint effect of both the cascade and the dissipation. In fact, we expect this time variation to be very small since the data have been stored after the DNS had reached a statistically stationary regime. This is confirmed on Fig. \ref{BalanceFigureb} where both black curves are close to zero. We also check that the divergence term vanishes at all scales (purple curves). We therefore conclude that Eq. (\ref{KHMglobal}) shorten to the RHS being zero.

Let us, then, study the variations in scales of the terms constituting the RHS. The global contribution of nonlinear effects (blue curves on Fig. \ref{BalanceFigureb}) varies as expected, being small at both large and small scales, and of maximum amplitude in the inertial range around $\ell/L = 0.125$. This maximum takes a value close to the total energy dissipation rate $\epsilon^T$, confirming our interpretation of $\Pi^\pm_\ell$ as describing scale-to-scale energy transfers. Besides, its contribution being negative, the cascade is indeed direct. The viscous term, plotted in green on Fig. \ref{BalanceFigureb}, is small in the injection and inertial ranges of scales, and becomes dominant over nonlinear effects at the end of the cascade, as usual. Finally, the forcing term (red curves) which describes the energy injected above scale $\ell$, is small at large scales and maximum in the inertial and dissipation ranges, balancing both energy dissipation and nonlinear transfers. Let us stress that $\langle\mathcal{F}^\pm_\ell\rangle_\mathcal{V}$ does not represent the global rate at which energy is injected at scale $\ell$. If this were the case, it would be maximum for $\ell/L \approx 1$ and zero at smaller $\ell$. Due to the filtering process which averages out fine details of the flow while keeping informations about large scales, $\langle\mathcal{F}^\pm_\ell\rangle_\mathcal{V}$ represents the energy injection at scales larger than $\ell$, explaining the variations of the red curves on Fig. \ref{BalanceFigureb}. Injection at scale $\ell$ would be $\langle\mathcal{I}^\pm_\ell\rangle_\mathcal{V} \coloneqq \epsilon^\pm - \langle\mathcal{F}^\pm_\ell\rangle_\mathcal{V}$, so that Eq. (\ref{KHMglobal}) becomes

\begin{equation}
\epsilon^\pm = \langle\Pi^\pm_\ell\rangle_\mathcal{V} + \langle\mathcal{D}^\pm_{\nu,\ell}\rangle_\mathcal{V} + \langle\mathcal{I}^\pm_\ell\rangle_\mathcal{V},
\end{equation}

at all $\ell$.

We now turn to the KHM relation in its local form. Out of the $1024^3$ available points, we have verified the local balance at several locations to check that it holds well. A typical example is given on Fig. \ref{BalanceFigurec} to \ref{BalanceFiguref}, where we show the LHS (solid lines) and RHS (dashed lines) of Eq. (\ref{localKHM}) as a function of scale at one point, together with the local contribution of each five terms constituting the balance. We observe that the local KHM relations hold well, thus providing numerical evidence that the local approach presented in Sec. \ref{LocalApproach} is valid in turbulent MHD data. This also confirms that the only processes breaking the local conservation of energy are injection, dissipation, and scale-to-scale transfers. Let us note, however, that the balance does not hold as well as what is shown in Fig. \ref{BalanceFigurec} and \ref{BalanceFiguree} for all points. Indeed, there exist many points where a larger discrepancy can be observed between the LHS and RHS of Eq. (\ref{localKHM}), especially at small scales. This can be explained by the fact that when $\ell$ is too small (typically of the order of a few grid steps), local averages are performed over few number points, so that they may not be well converged. For larger $\ell$ this discrepancy generally disappears.

Investigating further in details the local KHM relation, we can study the local variations of each five terms as a function of $\ell$. This is what is shown on Fig. \ref{BalanceFigured} and \ref{BalanceFiguref}, where we observe that the variations of all five terms are very different from Fig. \ref{BalanceFigureb}. Locally, the amplitude of these terms may be orders of magnitude larger than their space averages, and even change sign. Let us first focus on Fig. \ref{BalanceFigured}. We observe  that the local time variation of $e^+_\ell$ (black curve) is much stronger than its global average, and increases with decreasing scales. In the same way, we see that the divergence term (purple curve) is nonzero, and is one of the three dominant terms in the local balance throughout the available range of scales. It is strongly positive (compared to $\epsilon^+$) at large scales, and strongly negative in the dissipation range. Local scale-to-scale energy transfers (blue curve) also locally change sign in the inertial range, have a strong amplitude, and fall to zero at very small and very large scales. Finally, the dissipation and forcing terms (respectively green and red curves) are negligible compared to the other three. We can therefore interpret the results presented on Fig. \ref{BalanceFigured} as follows: in the inertial range, where the $\Pi^+_\ell$ changes sign, $\bm z^+$ locally accumulates energy due to a converging spatial energy flux (negative divergence). This flux, then, plays the role of a forcing, providing energy which feeds two cascades to both large and small scales. At large scales, energy is provided by this local inverse cascade, and leaves the area under consideration via a spatial energy flux directed outwards. However, the rate of the cascade is larger than the rate at which energy is removed, thus leading to an accumulation of energy in $\bm z^+$. At small scales, the roles of the spatial flux and the nonlinear interactions are reversed. Energy is brought by the local flux and removed by the cascade. Again, energy is brought faster than it is removed, the local cascade vanishing at small scales. This again leads to a local accumulation of energy, the local KHM relation (\ref{localKHM}) being preserved.

If we now turn to Fig. \ref{BalanceFiguref}, the interpretation is slightly different since, roughly, the signs are reversed (same color code as in Fig. \ref{BalanceFigured}) . As a consequence, we see that the energy $e^-_\ell$  locally contained above scale $\ell$ in $\bm z^-$ is decreasing in time, except for large $\ell$. At large scales, we have a local direct cascade of energy together with an incoming spatial energy flux. The viscous and forcing terms being once again negligible, $e^-_\ell$ is increasing at large scales since the incoming rate is stronger than the cascade rate. This competition between nonlinear transfers and the incoming flux continues through the inertial range, where the cascade rate becomes slightly stronger than the flux, therefore leading to a decreasing $e^-_\ell$. At some point in scale the cascade and the flux change sign, with an outgoing flux always stronger than the inverse cascade. This explains the local decrease of $e^-_\ell$ at small scales.

In conclusion, we have checked that the local KHM relation (\ref{localKHM}) is locally satisfied, and highlighted that its constituting terms may exhibit much stronger amplitudes than their space averages, as could be expected. Moreover, the local behaviour of one same term may be very different  for both balances, and local cross-scale transfers may change sign as a function of $\ell$.


\subsection{Local organization of scale-to-scale energy transfers}
\label{localOrg}

\begin{figure}	
	\centering
	\begin{subfigure}[t]{8.5cm}
		\caption{}
		\label{TransfLoca}
		\begin{overpic}[width=9.1cm]{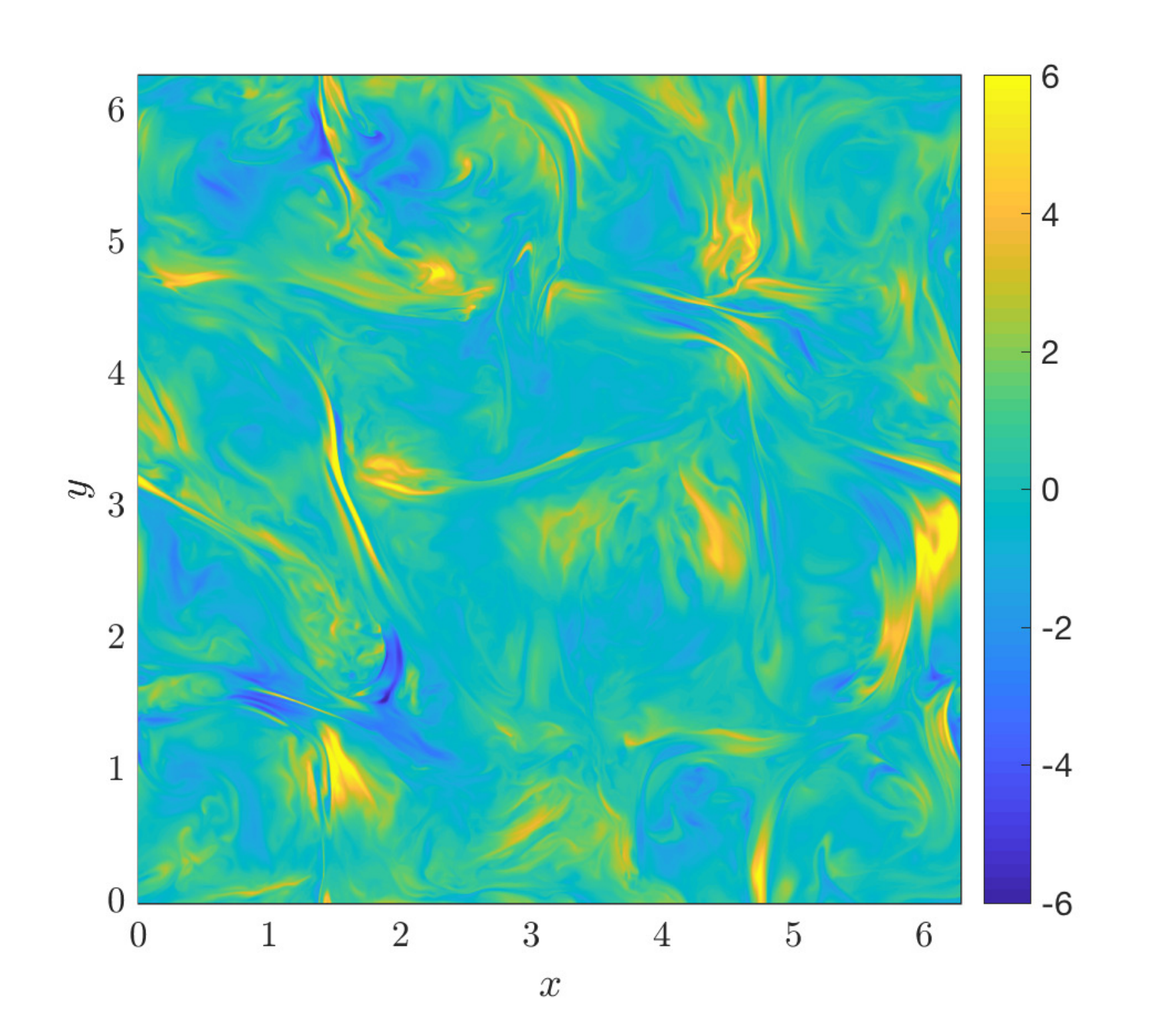}
       \put(55,83){\large $\ell/L = 0.996$}
       \end{overpic}			
	\end{subfigure}
	\quad
	\begin{subfigure}[t]{8.5cm}
		\caption{}
		\label{TransfLocb}
		\begin{overpic}[width=9.1cm]{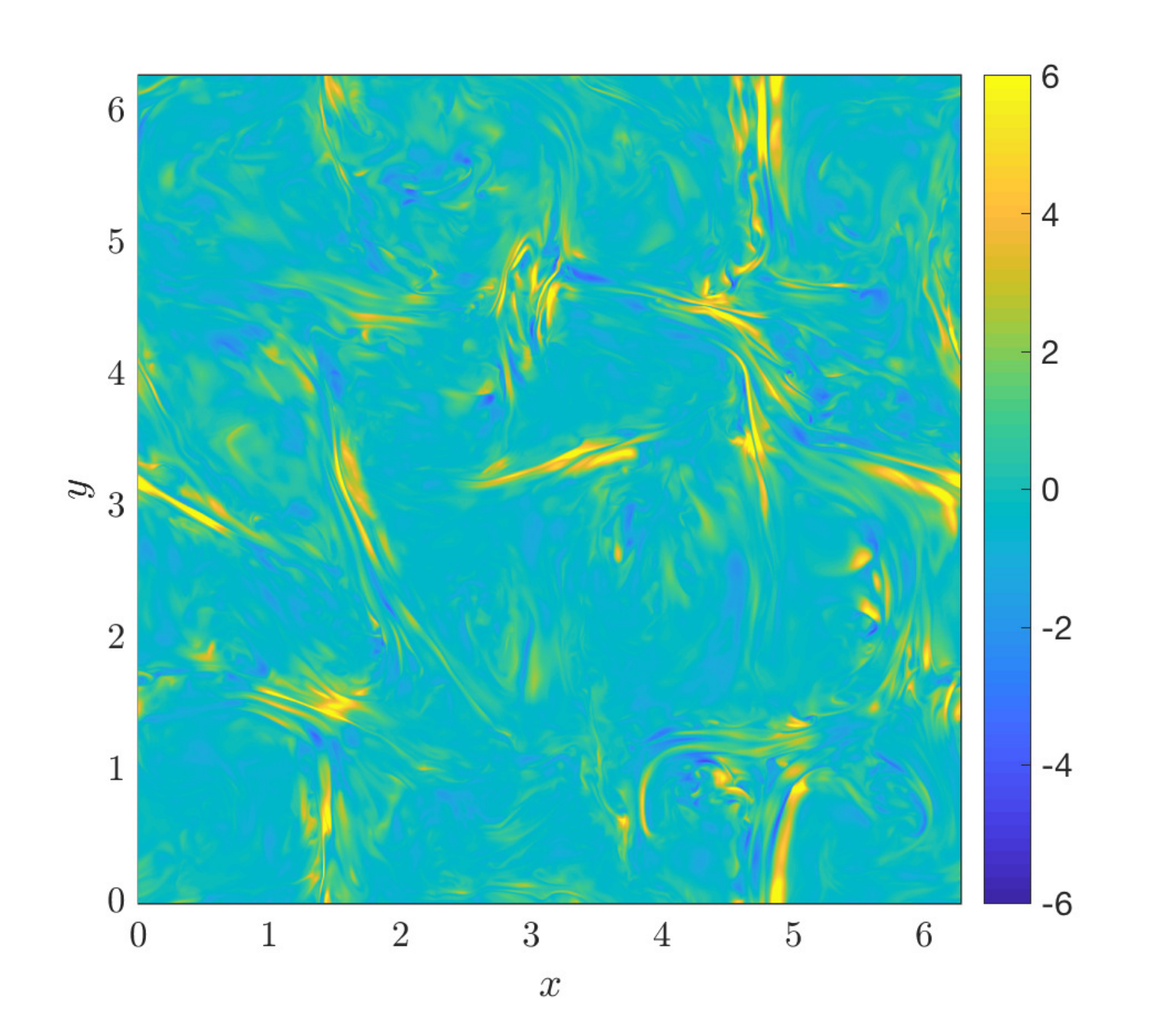}
		 \put(15,83){\large $\ell/L = 0.125$}
       \end{overpic}
	\end{subfigure}
		\begin{subfigure}[t]{8.5cm}
		\caption{}
		\label{TransfLocc}
		\begin{overpic}[width=9.1cm]{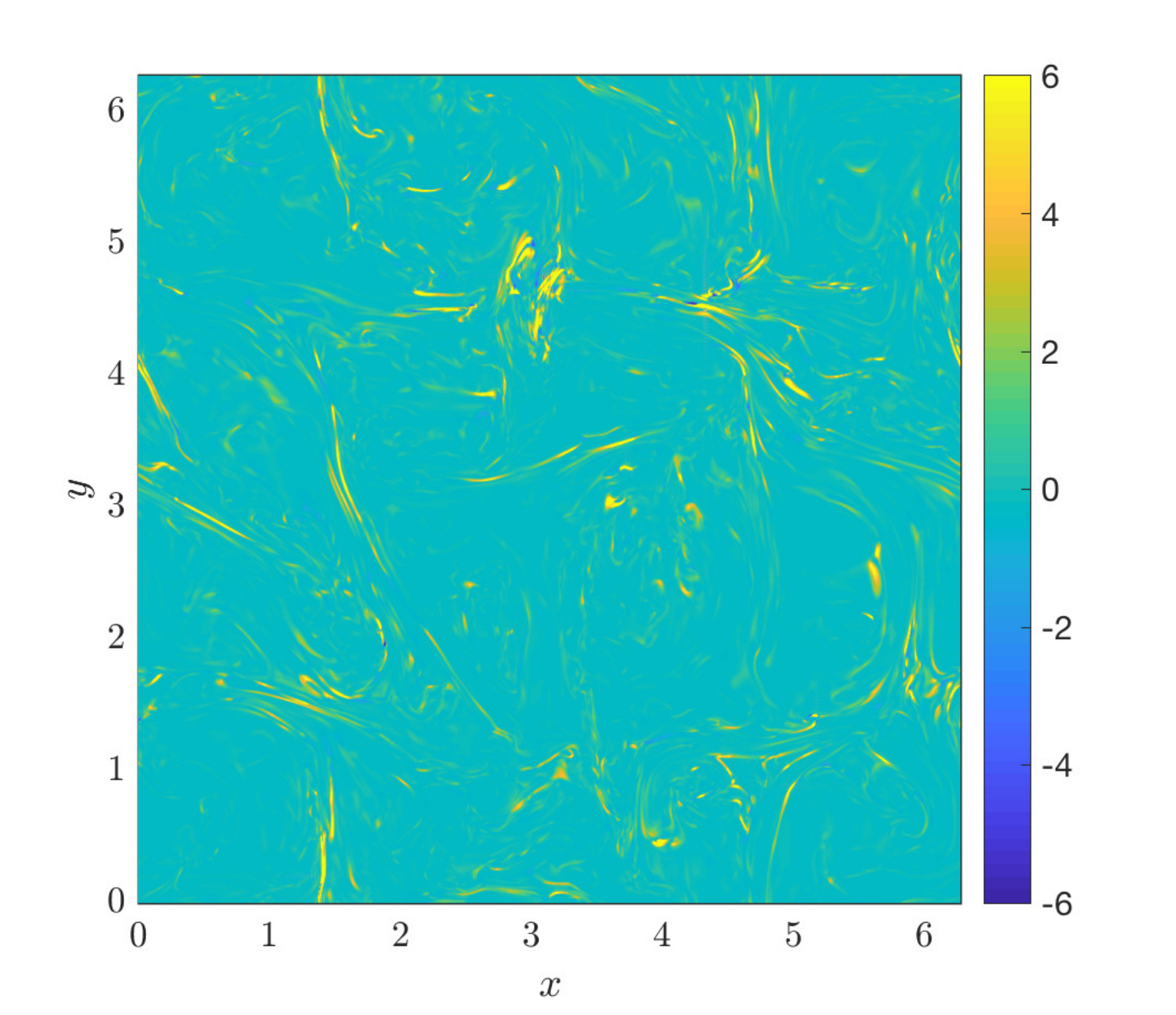}
		 \put(55,83){\large $\ell/L = 0.0078$}
       \end{overpic}		
	\end{subfigure}
	\caption{Typical slices of $\Pi^T_\ell$ in the same plane as Fig. \ref{Fields} at three different scales. (a) close to the injection scale, (b) in the inertial range, (c) in the dissipative range. The mean and variance of $\Pi^T_\ell$ have been set to zero and unity respectively at all three scales for better comparison. We observe that local transfers are organized into structures existing over a wide range of scales.}
	\label{TransfLoc}
\end{figure}

In the rest of the paper, we are going to focus on transfers of total energy (kinetic plus magnetic) through scales. Locally, these transfers are described by the quantity

\begin{equation}
\Pi^T_\ell \coloneqq  \frac{\Pi^+_\ell + \Pi^-_\ell}{2}.
\end{equation}

We show on Fig. \ref{TransfLoc} typical slices of $\Pi^T_\ell$ in the $\left(xy\right)$ plane. These slices are displayed at three different scales: $\ell/L = 0.996$ close to the injection scale (Fig. \ref{TransfLoca}), $\ell/L = 0.125$ in the inertial range (Fig. \ref{TransfLocb}), and $\ell/L = 0.0078$ in the dissipative range, approximately three times larger than both hydrodynamic and magnetic Kolmogorov length scales (Fig. \ref{TransfLocc}). Since the statistics of energy transfers (which are discussed in Sec. \ref{PiStat}) depend on $\ell$, we have normalized $\Pi^T_\ell$ so that it has zero space mean and unit variance at the three displayed scales, which allows for a better comparison.

First of all, we observe that energy transfers fluctuate in space and are organized into local structures characterized by strong magnitudes in a background of weak (or average) transfers. Qualitatively, we call "structures" events of strong cross-scale transfers correlated over a large number of neighbouring grid points (typically $10^4-10^5$ at inertial scales). As we see from Fig. \ref{TransfLoc}, these structures exist over a wide range of scales, some of them covering all available $\ell$. Moreover, we observe that they get more localized as $\ell$ is decreased, in agreement with the results of Sec. \ref{CheckBalance} where we saw that viscous effects become globally dominant at small scales. As a consequence, they seem to occupy a smaller fraction of space at smaller scales, as reported by previous MHD and pure hydrodynamic studies \cite{Saw2016, Kuzzay2017, Camporeale2018}. It is interesting to note that nonlinear effects do not vanish close to Kolmogorov scale and can be locally very strong. Indeed, Fig. \ref{TransfLocc} shows local events which deviate more than six times the standard deviation from their mean. Such strong events have also been observed in hydrodynamic turbulence in the framework of von Kármán flows \cite{Saw2016}. Similar results at both ion and electron scales have been reported in maps of cross-scale kinetic energy transfers as well, using a LES approach, in the simulation data of a fully kinetic particle-in-cell simulation \cite{Yang2017}. This highlights the local nature of the mechanisms responsible for the cascade, and shows the importance of probing very small scales to understand turbulent energy dissipation. However, let us insist that caution should be taken when discussing the physics of the dissipative range from our study since local averages may be badly converged due to the small number of grid points contained contained in the ball of size $\ell$. Finally, let us observe that these structures can be either positive or negative, meaning that energy can be locally transferred from large to small scales or backscattered, with an overall transfer to small scales as evidenced in Fig. \ref{BalanceFigureb}.

\subsection{Statistics of local energy transfers}
\label{PiStat}

\begin{figure}	
	\centering
	\begin{subfigure}[t]{8.5cm}
		\caption{}
		\label{muT}
		\includegraphics[width=9.1cm]{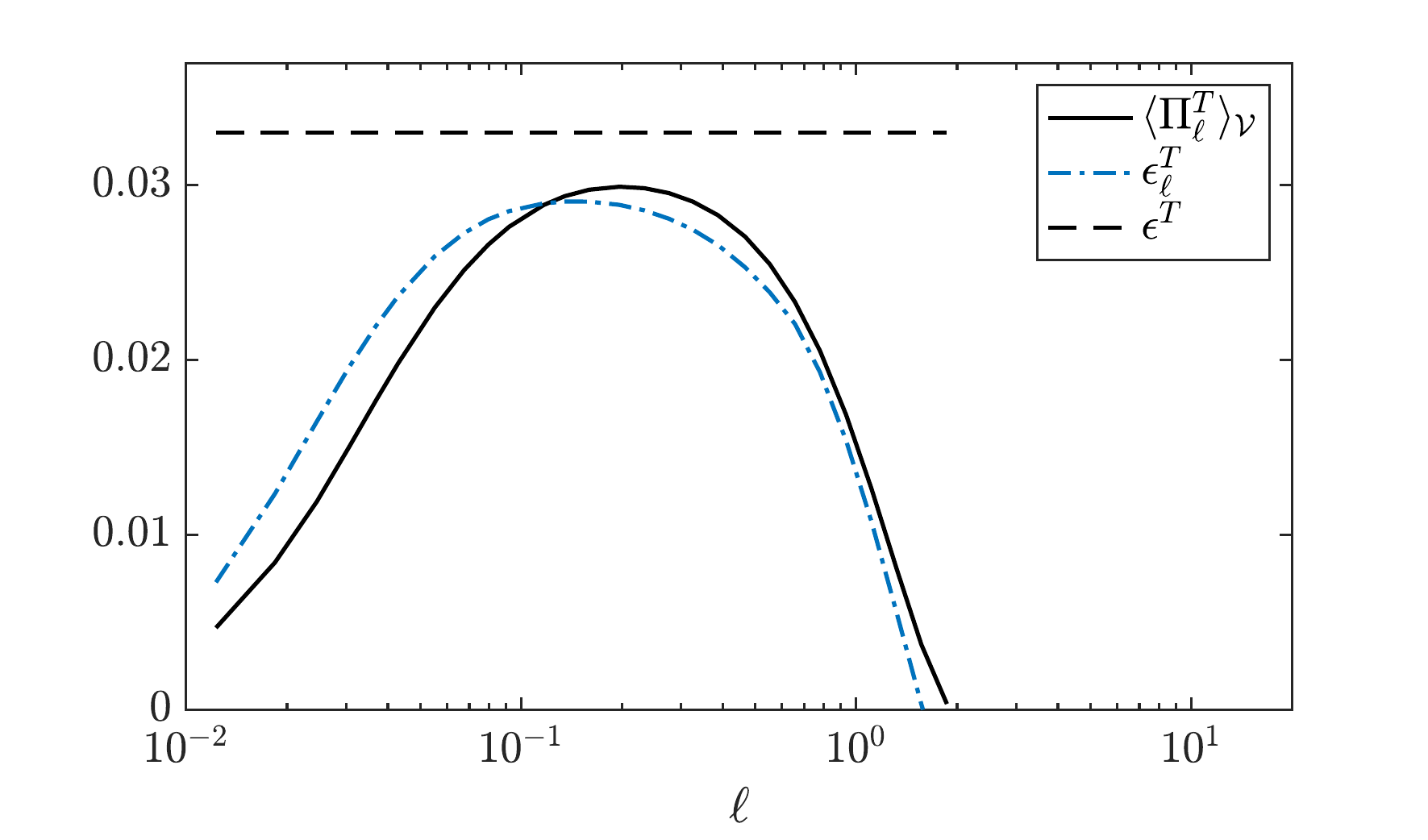}			
	\end{subfigure}
	\quad
	\begin{subfigure}[t]{8.5cm}
		\caption{}
		\label{sigmaT}
		\includegraphics[width=9.1cm]{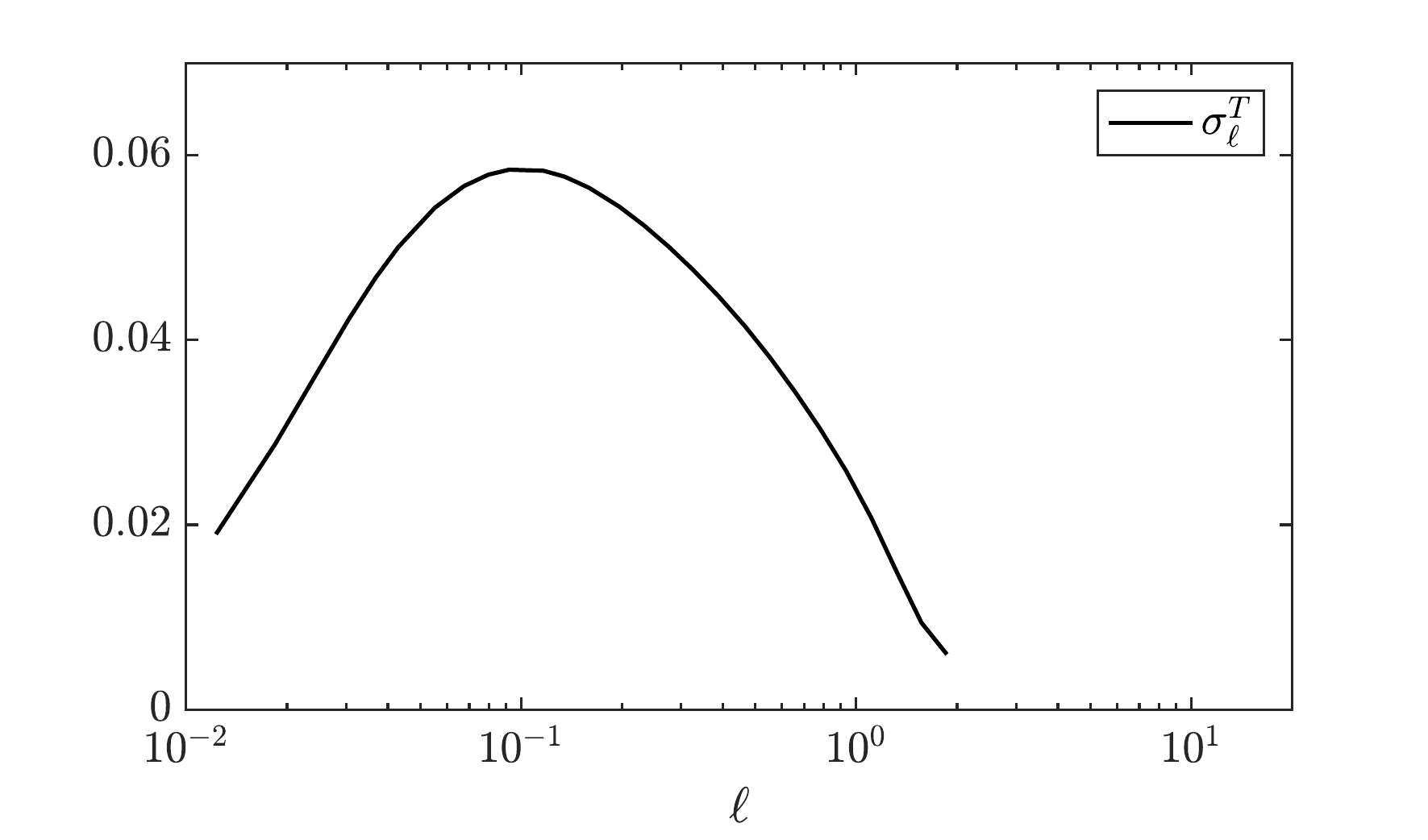}	
	\end{subfigure}
	\begin{subfigure}[t]{8.5cm}
		\caption{}
		\label{PDF}
		\includegraphics[width=9.1cm]{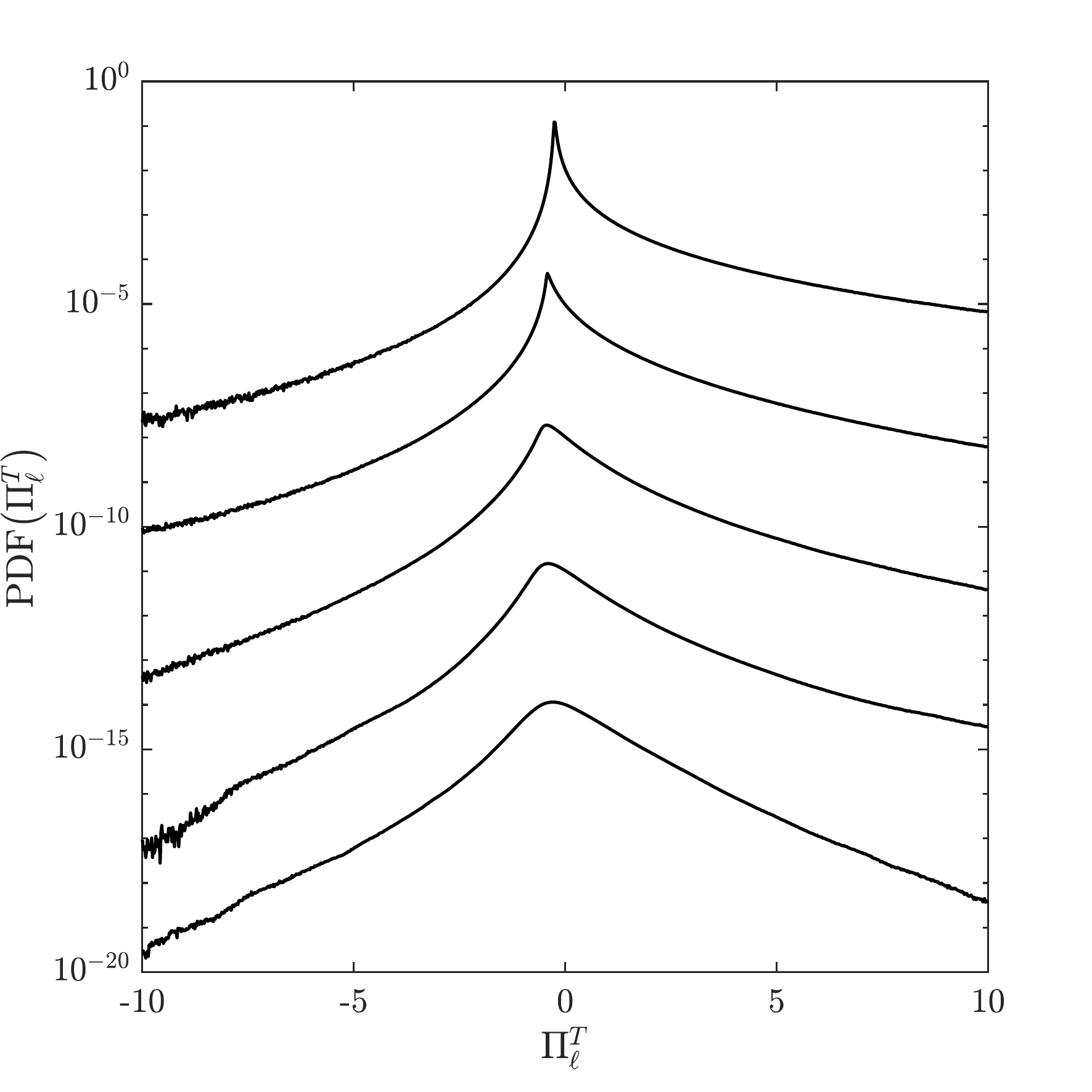}	
	\end{subfigure}
	\caption{(a) Global scale-to-scale transfers as a function of scale $\ell$. (b) Standard deviation $\sigma^T_\ell$ of $\Pi^T_\ell$ as a function of scale $\ell$. (c) Probability distribution functions of the local scale-to-scale transfers $\Pi^T_\ell$. Represented scales (from top to bottom): $\ell/L$ = 0.0078, 0.0352, 0.125, 0.352, 0.996. All curves have been arbitrarily vertically shifted for the sake of clarity, and their mean and variance have been respectively set to zero and unity.}
	\label{PiStatFig}
\end{figure}

We have seen in Sec. \ref{localOrg} that $\Pi^T_\ell$ is a fluctuating quantity, the statistics of which depend on scale $\ell$. Here, we are going to investigate these statistical properties. Fig. \ref{PiStatFig} displays the spatial mean (Fig. \ref{muT}), standard deviation (Fig. \ref{sigmaT}), and probability density functions (Fig. \ref{PDF}) of $\Pi^T_\ell$ as a function of scales. From Sec. \ref{CheckBalance}, we know that $\Pi^+_\ell$, $\Pi^-_\ell$, and therefore $\Pi^T_\ell$ have approximately the same averages at all scales (solid black curve on Fig. \ref{muT}). Fig. \ref{muT} also displays the variations of

\begin{equation}
\epsilon^T_\ell \coloneqq \frac{\epsilon^+_\ell + \epsilon^-_\ell}{2},
\end{equation}

computed from third order structure functions (dash-dotted blue curve), which allows for the comparison of our approach with Politano-Pouquet's 4/3-law. Indeed, in the inertial range where homogeneity and isotropy are expected to be recovered, space averages are equivalent to ensemble averages so that $\langle\Pi^T_\ell\rangle_\mathcal{V}$ should compare well with $\epsilon^T_\ell$. This is what we observe on Fig. \ref{muT}, where both quantities behave essentially the same despite slight discrepancies. For example, the maximum of $\epsilon^T_\ell$ is slightly smaller than the one of $\langle\Pi^T_\ell\rangle_\mathcal{V}$, and is reached at $\ell/L = 0.102$ instead of $\ell/L = 0.125$. There are mainly two reasons for these discrepancies. First, in the computation of $\langle\Pi^T_\ell\rangle_\mathcal{V}$, we perform local averages of $\delta_r z^\mp_j \vert \delta_r \bm z^\pm\vert^2$ over scales weighted by $\left(\partial_j G\right)_\ell$ prior to summing over the whole box. These local averages make $\Pi^T_\ell$ smoother than the raw quantity $\delta_r z^\mp_j \vert \delta_r \bm z^\pm\vert^2$, and slightly different results are therefore expected after space-averaging. Second, the flow is inhomogeneous and anisotropic due to the large-scale forcing. We observe in the data that homogeneity and isotropy are not entirely recovered at smaller scales so that some anisotropy and inhomogeneity persist at inertial scales. This is due to the Reynolds numbers not being high enough to create a well-defined inertial range decoupled from large-scale effects. Since the 4/3-law (\ref{4/3MHD}) assumes both homogeneity and isotropy, $\epsilon^T_\ell$ does not to exactly provide a measure of the energy cascade rate in the inertial range. This concretely emphasizes the power of the local approach which is free from homogeneity and isotropy assumptions.

Finally, let us note that finite Reynolds number effects are also responsible for the fact that both estimates of the energy cascade rate are smaller than the energy dissipation rate $\epsilon^T$ (horizontal dashed line on Fig. \ref{muT}). Indeed, relations (\ref{4/3MHD}) and (\ref{4/3local}) are asymptotic laws, expected to be valid in the limit of infinite Reynolds numbers, or at least in a well-defined inertial range, far from the injection and dissipation scales. At finite Reynolds numbers, it has been shown that effects from large and viscous scales are not negligible at inertial scales, so that the 4/3-law is not exactly satisfied even when the flow is homogeneous and isotropic. However, when the Reynolds number is increased, a decrease of the discrepancies from asymptotic laws is observed, together with an extension of the inertial range \cite{Qian1997, Gagne2004, Bos2012}. Nonetheless, we can conclude our discussion of Fig. \ref{muT} by emphasizing that our approach is in very good agreement with the usual statistical one. \\

The standard deviation of $\Pi^T_\ell$ is defined as

\begin{equation}
\sigma^T_\ell \coloneqq \sqrt{\langle{\Pi^T_\ell}^2\rangle_\mathcal{V} - \langle\Pi^T_\ell\rangle_\mathcal{V}^2}.
\end{equation}

Its variations with scales are displayed on Fig. \ref{sigmaT}, where we observe that as for the mean, it is small at large scales, reaches its maximum around $\ell/L = 0.0586$ in the inertial range, and then decreases at small scales. This confirms that the highest magnitudes of scale-to-scale transfers are most likely to be reached at inertial scales. Moreover, we see that the standard deviation is between two and three times larger than the mean for all $\ell$. This indicates that the magnitude of scale-to-scale transfers tend not to be clustered around their mean, but instead deviate significantly. The way in which they are spread is given by plotting the probability distribution functions (PDFs) of $\Pi^T_\ell$ at various scales. These PDFs are displayed on Fig. \ref{PDF} where their mean is set to zero, their variance to unity, and they have been arbitrarily shifted vertically for the sake of clarity. First, we observe that they are widely spread out with very large tails, exhibiting the existence of events where

\begin{equation}
\vert \Pi^T_\ell - \langle\Pi^T_\ell\rangle_\mathcal{V}\vert \geqslant 10\times\sigma^T_\ell.
\end{equation}

The existence of such large tails explains the large standard deviations on Fig. \ref{sigmaT}. Moreover, we see that the PDFs undergo a continuous shape deformation as $\ell$ is decreased. At injection scales, the PDF is almost symmetric while it becomes positively skewed and strongly peaked around small positive values in the inertial and dissipative range. These peaks are shifted to negative values of $\Pi^T_\ell$ on Fig. \ref{PDF} because we display the PDFs with zero mean. Note that these results differ from those reported in \cite{Coburn2014, Coburn2015} for the inertial range of solar wind turbulence. In their study, the authors compute the PDFs of third-order moments from time series of Advanced Composition Explorer. Although our results confirm that the standard deviations of the PDFs are large compared to their means, and observe backscatter as well, the authors of \cite{Coburn2014, Coburn2015} find that their PDFs resemble Gaussians with a small skewness. Even though we refrain from quantifying the skewness and kurtosis of the PDFs displayed in Fig. \ref{PDF} as we doubt that statistics of order nine and twelve would be well converged, the statistics we observe here are clearly not gaussian and further studies are needed in order to understand the origin of these different results. Nonetheless, our results show good agreement with previous studies from hydrodynamics \cite{Saw2016, Kuzzay2017}. Finally, it can be understood from our discussion in Sec. \ref{localOrg} that the structures we observe on Fig. \ref{TransfLoc} are clusters of events constituting the tails of the PDFs, while their overall positive magnitudes are explained by the skewness of the PDFs. However, the PDFs do not provide any further explanation concerning the apparent decrease of the amount of space these structures occupy as we investigate smaller scales.

\begin{figure}	
	\centering
	\begin{subfigure}[t]{8.5cm}
		\caption{}
		\label{SpaceFilling}
		\includegraphics[width=9.1cm]{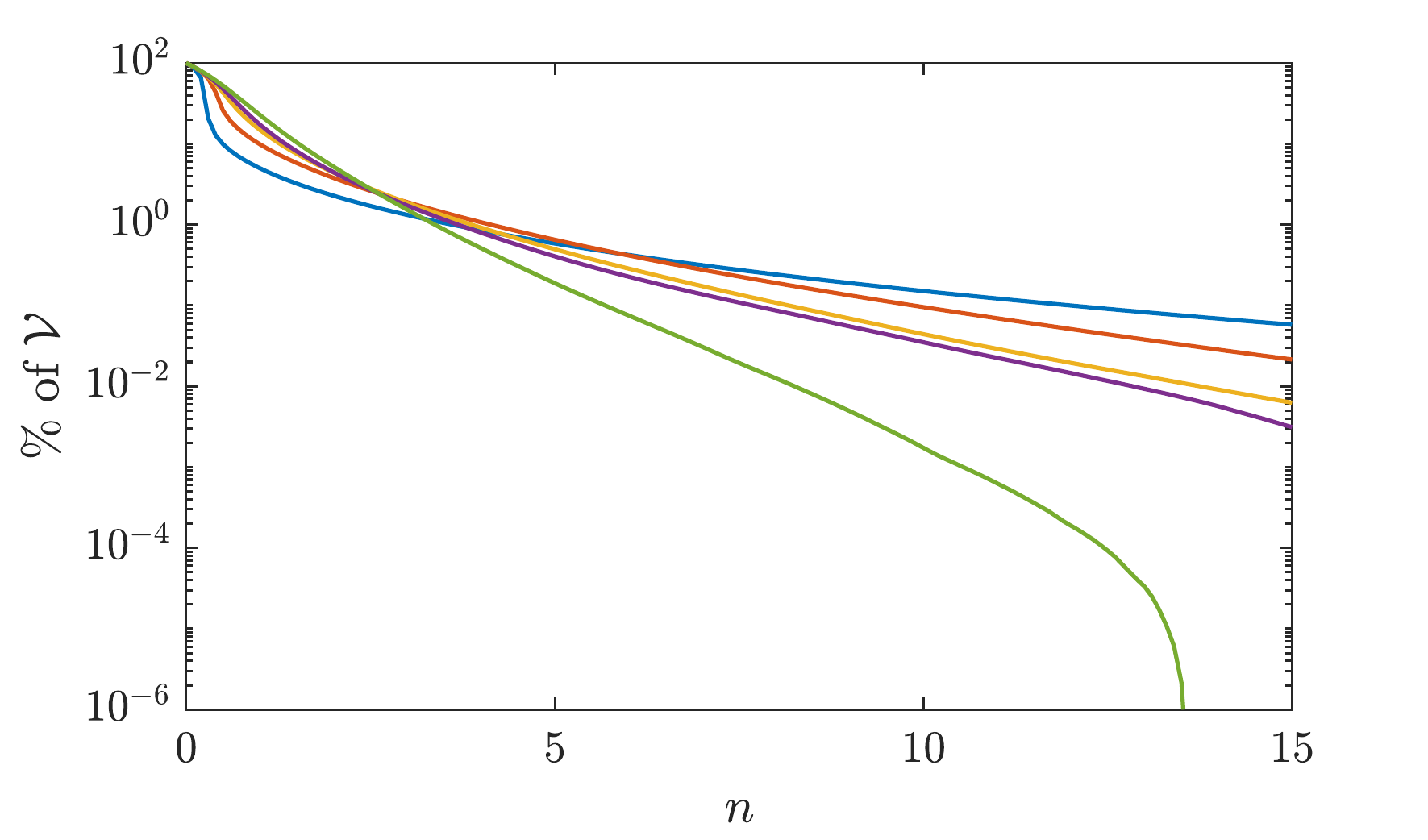}			
	\end{subfigure}
	\quad
	\begin{subfigure}[t]{8.5cm}
		\caption{}
		\label{TransFilling}
		\includegraphics[width=9.1cm]{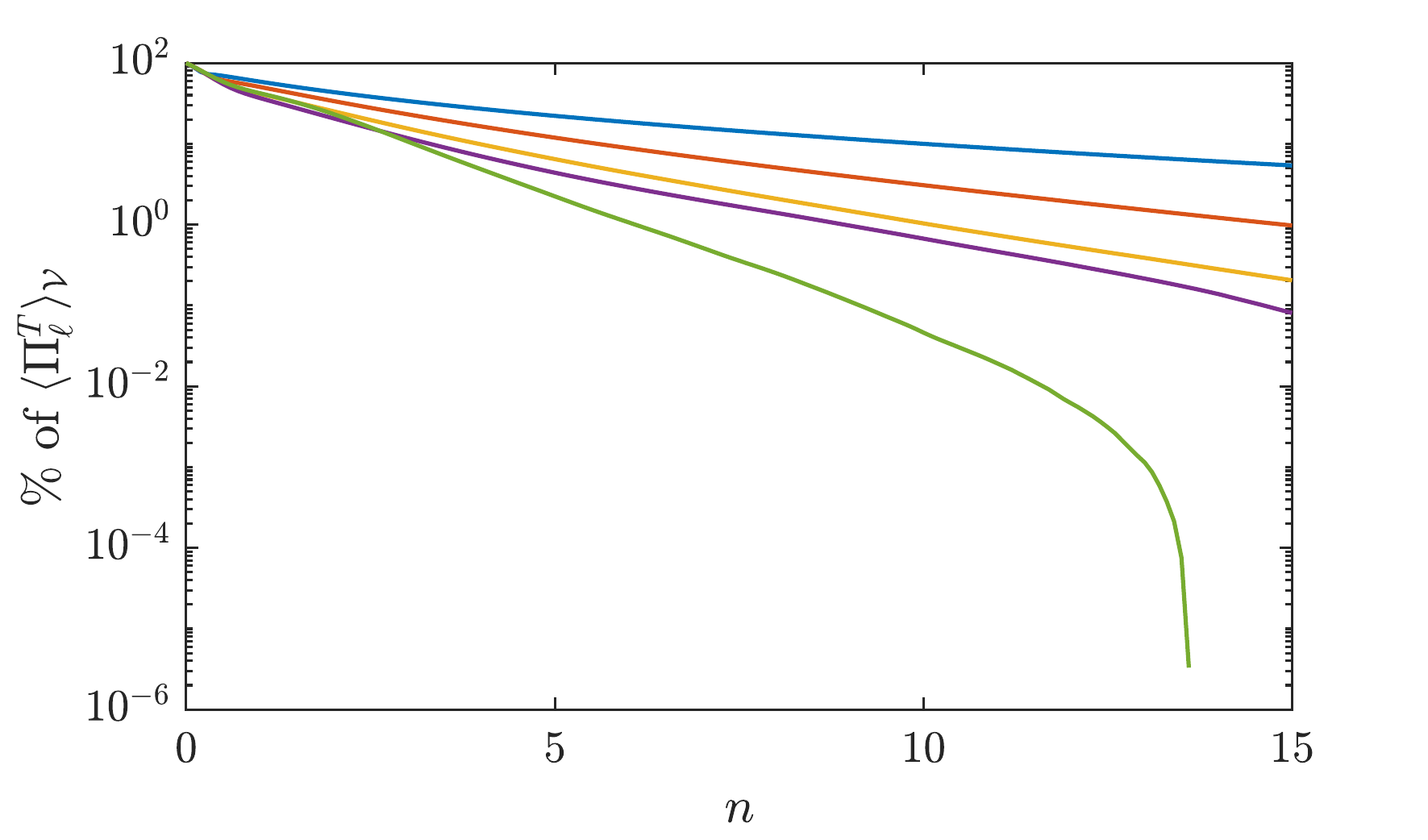}	
	\end{subfigure}
	\begin{subfigure}[t]{8.5cm}
		\caption{}
		\label{TransSpace}
		\begin{overpic}[width=9.1cm]{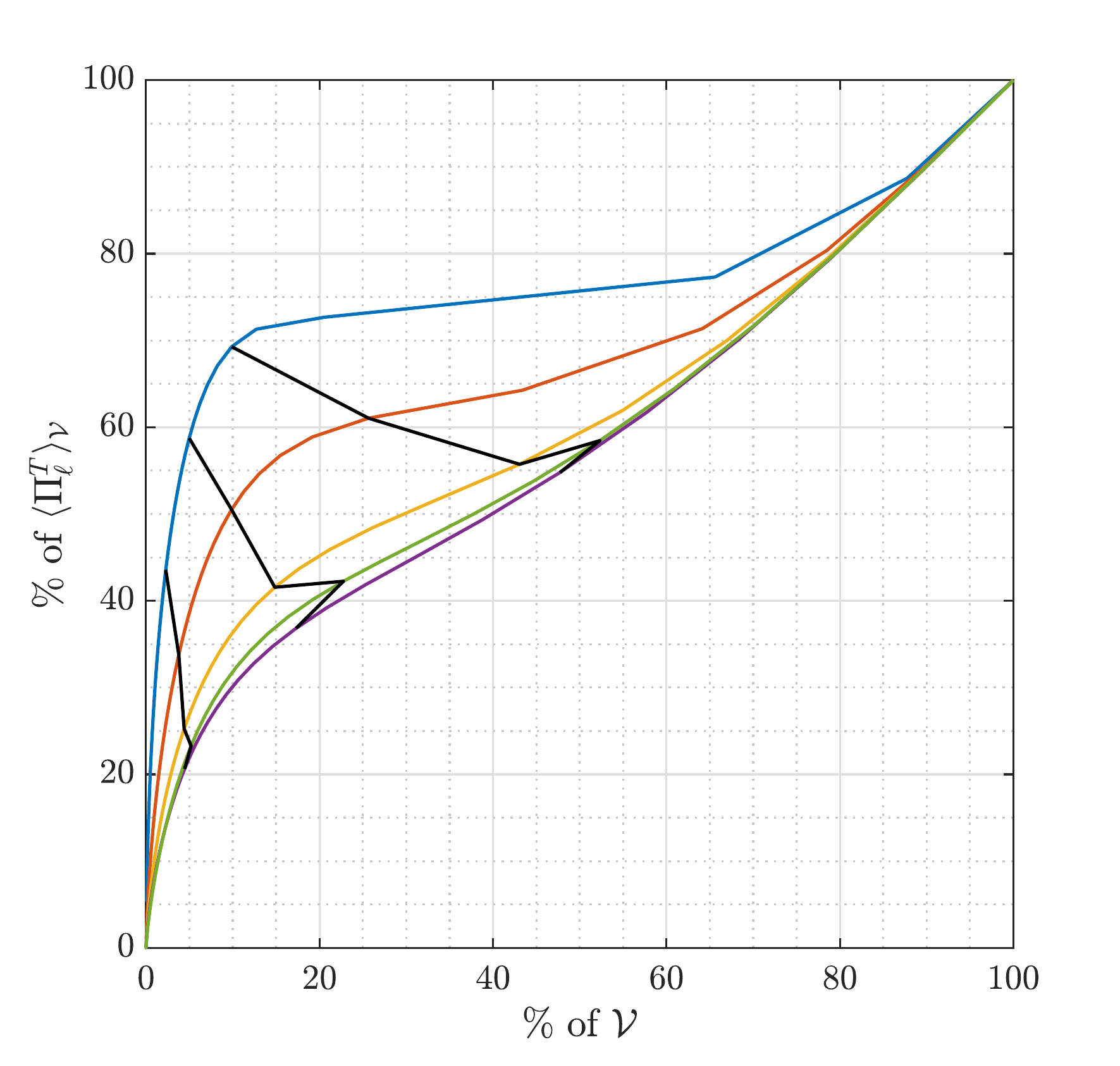}
       \put(58,54){$n = 0.5$}
       \put(33,39){$n = 1$}
       \put(21,27){$n = 2$}
       \end{overpic}
	\end{subfigure}
	\caption{(a) Estimate of the fraction of space occupied by events which deviate more than $n\times\sigma^T_\ell$ from their mean as a function of $n$. (b) Estimate of the contribution to global scale-to-scale transfers of events which deviate more than $n\times\sigma^T_\ell$ from their mean as a function of $n$. (c) Fraction of space in which a given fraction of the global scale-to-scale transfers is contained. Represented scales: $\ell/L$ = 0.0078 (blue), 0.0352 (red), 0.125 (yellow), 0.352 (purple), 0.996 (green).}
	\label{FillingFactor}
\end{figure}

This can be studied by computing the filling factor of events which deviate more than some given threshold from their mean. For this purpose, we search in our data for every points where

\begin{equation}
\vert \Pi^T_\ell - \langle\Pi^T_\ell\rangle_\mathcal{V}\vert \geqslant n\times\sigma^T_\ell,
\end{equation}

for various values of $n$ and $\ell$. Note that with this criterion, we focus on points belonging to the tails of the PDFs, and which therefore constitute structures if $n$ is chosen not too small. The fraction of points we detect will provide us with an estimate of the fraction of space occupied by these structures, and the total sum of $\Pi^T_\ell$ at these points will yield an estimate for their contribution to the energy cascade rate. The results are reported on Fig. \ref{FillingFactor}.

Fig. \ref{SpaceFilling} displays the space filling factor of these points as a function of $n$ and $\ell$. First of all, we observe that it decreases very rapidly as $n$ is increased. For example, it drops to $10\%$ for $n \approx 1.5$ at injection and inertial scales (green and yellow curves respectively), and for $n \approx 0.5$ close to Kolmogorov scale (blue curve). Second, we see that for $n\leqslant 2$ the filling factor increases with scale if $n$ is fixed, whereas it decreases for $n \geqslant 6$. The faster decrease at smaller scales for $n\leqslant 2$ is in agreement with our observation of Fig. \ref{TransfLoc} that structures get more localized as $\ell$ is decreased. However, the reversal observed for larger values of $n$ may be surprising. Indeed, this means that the most statistically extreme events are more abundant at smaller scales. This seems counter intuitive and indicates that locally strong nonlinear effects exist and should be studied in more details, even in the dissipative range.

This remark is confirmed by Fig. \ref{TransFilling}, where the contribution of the same events to the overall transfer rate $\langle\Pi^T_\ell\rangle_\mathcal{V}$ is plotted. Again, this quantity is a decreasing function of $n$. However, contrary to the space filling factor, it increases with decreasing scale for almost all values of $n$. This shows that there is an increasing contribution to $\langle\Pi^T_\ell\rangle_\mathcal{V}$ of the events in the tails of the PDFs at smaller scales. For instance, events with $n \geqslant 10$ contribute to $0.05\%$ of the cascade rate at large scale (green curve), $1\%$ at inertial scale (yellow curve), and $10\%$ close to Kolmogorov scale. It is quite remarkable that such extreme events contribute so largely to the global transfer rate at scales where viscous forces dominate the dynamics. Note also that at this scale we have $\langle\Pi^T_\ell\rangle_\mathcal{V} = 0.0047$ while $\sigma^T_\ell = 0.019$. This means that events which are characterized by $n \geqslant 10$ have a magnitude larger than $5.7\epsilon^T$ in absolute value. The existence of strong nonlinear effect close to Kolmogorov scale was observed in experimental hydrodynamic flows \cite{Saw2016}, and their study is left for future work in the MHD case. To summarize, the results from Fig. \ref{SpaceFilling} and Fig. \ref{TransFilling} show that the events which are far in the tails of the PDFs of $\Pi^T_\ell$, and which are organized into structures in Fig. \ref{TransfLoc}, become increasingly localized as $\ell$ is decreased, but contribute to an increasing part of the cascade rate. This result can be made more explicit by plotting the fraction of global transfers due to the events characterized by a certain value of $n$ as a function of the fraction of space they occupy, \textit{i.e.} the ordinate in Fig. \ref{TransFilling} as a function of the ordinate in Fig. \ref{SpaceFilling}. This plot is displayed in Fig. \ref{TransSpace}, each point of a curve corresponding to a different value of $n$. We can clearly see the increasing importance of localized structures to the cascade rate as $\ell$ decreases. For instance, events corresponding to $n \geqslant 1$ occupy roughly $20\%$ of space and contain $40\%$ of the global transfer rate at large scale (green curve), while they occupy $15\%$ and $5\%$ of space at inertial (yellow curve) and dissipative scales (blue curve), and contribute respectively to $40\%$ and $60\%$ of the global transfer rate. This confirms that structures of strong events get more localized for decreasing scales, while contributing to an increasing fraction of the global transfers. This also emphasizes the importance of local strong events of scale-to-scale transfers, even in the dissipative range, and the value of the local approach to the study of the turbulent cascade.

Finally, let us note that the results displayed in Fig. \ref{TransSpace} differ from those presented in \cite{Camporeale2018}. In their study, the authors find only a small difference of the repartition of scale-to-scale transfers in space as $\ell$ is varied, whereas we observe an increasing difference as we investigate smaller scales. Moreover, we find that these transfers are much more concentrated into smaller regions of space than what is seen in their results. However, these differences can be explained easily. As we noted in Sec. \ref{linkKHM}, their expression for the local scale-to-scale transfers differs from ours due to different definitions of the large-scale energy. More importantly, they consider the distribution in space of the absolute value of these transfers rather than their algebraic sum, which changes the estimation of the filling factors. Finally, their data come from a 2D Hall-MHD DNS with nonzero mean magnetic field, in which the local mechanisms leading to the energy cascade are expected to be different from the 3D MHD case without mean magnetic field.

\subsection{3D structures of energy transfers and space gradients}

\begin{figure}	
	\centering
	\begin{subfigure}[t]{8.5cm}
		\caption{}
		\label{3Dtout}
		\includegraphics[width=9.1cm]{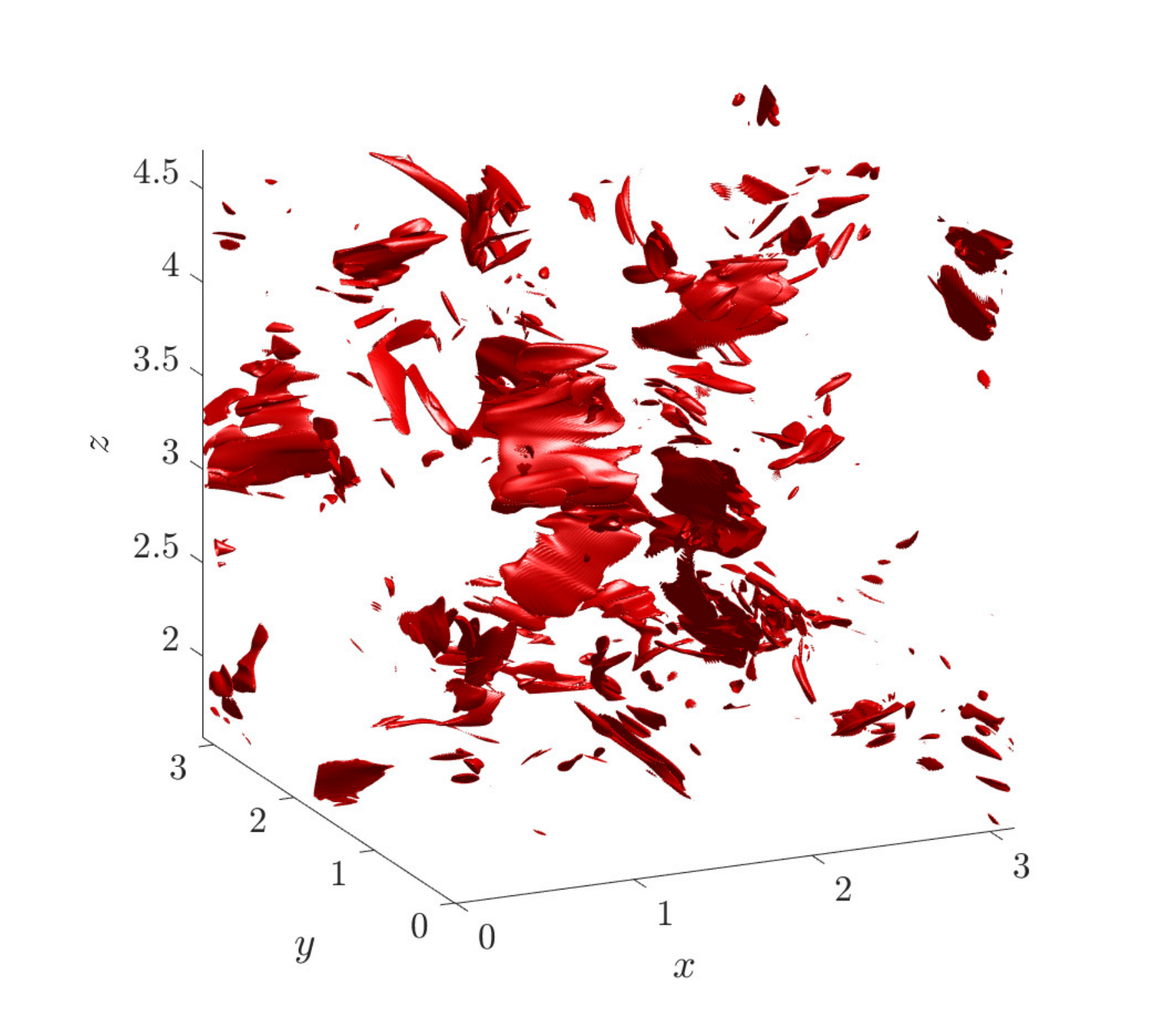}			
	\end{subfigure}
	\quad
	\begin{subfigure}[t]{8.5cm}
		\caption{}
		\label{nappe}
		\includegraphics[width=9.1cm]{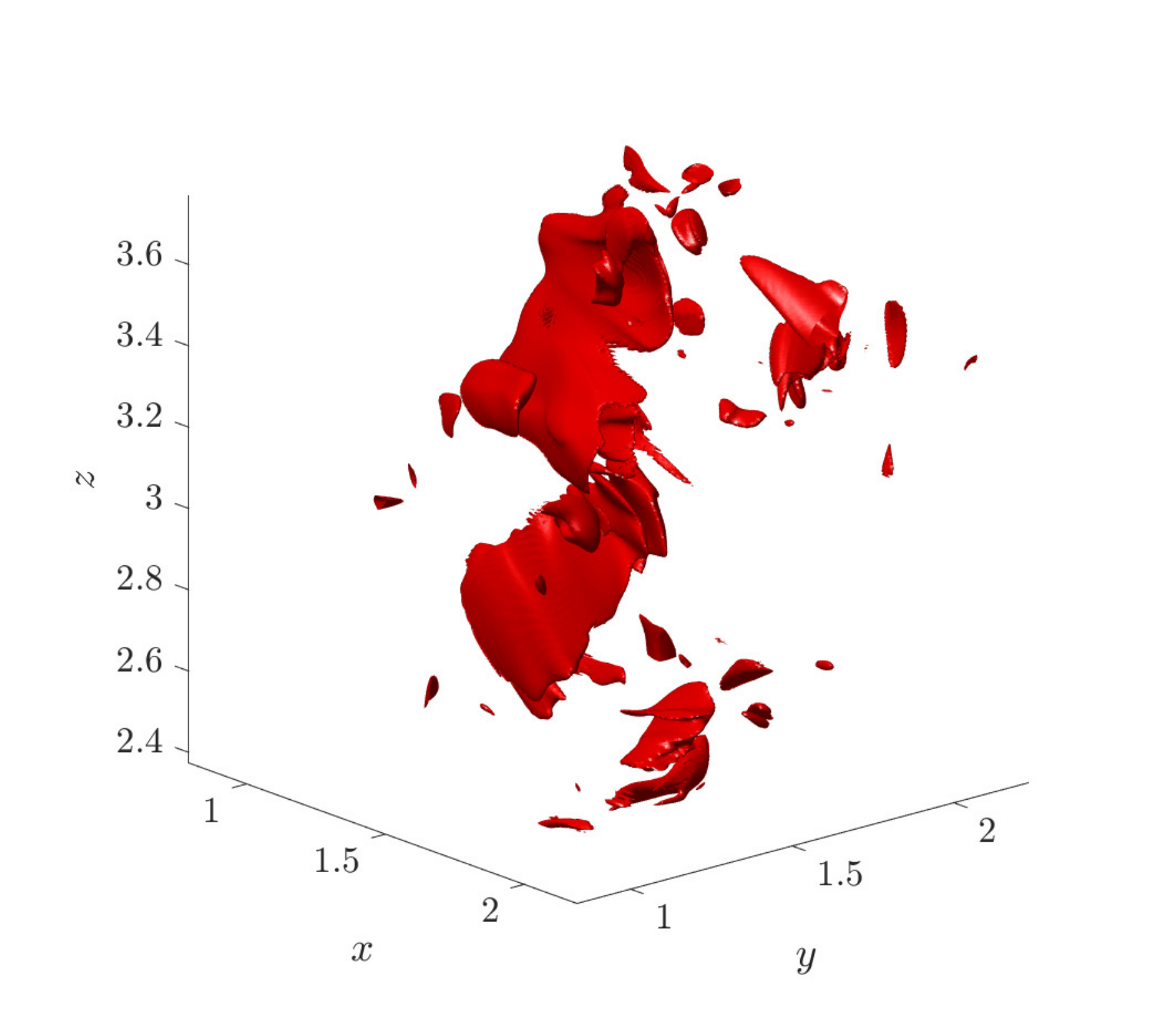}	
	\end{subfigure}
	\begin{subfigure}[t]{8.5cm}
		\caption{}
		\label{fils}
		\includegraphics[width=9.1cm]{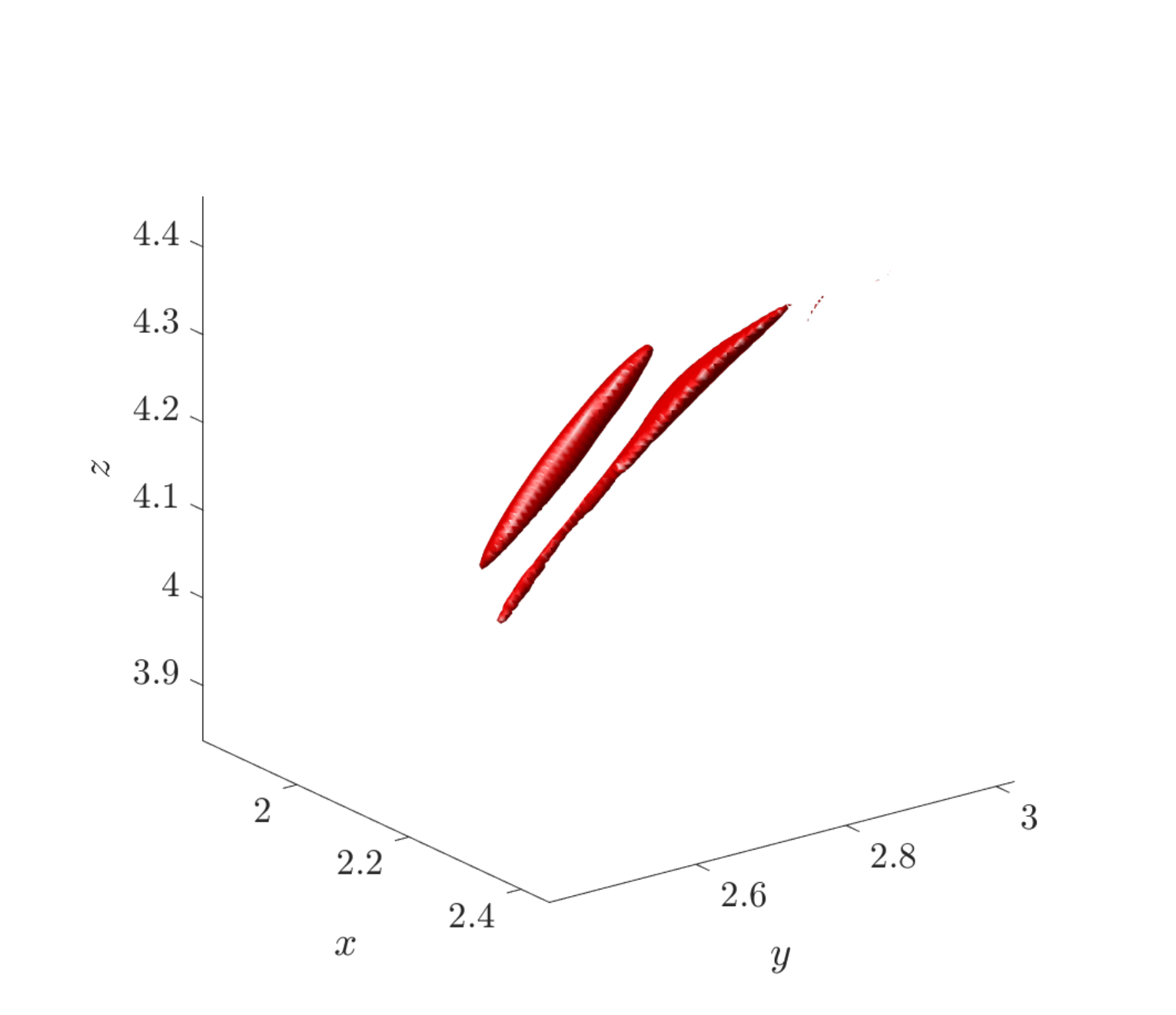}	
	\end{subfigure}
	\caption{Three-dimensional organization of cross-scale transfer structures at $\ell/L = 0.125$ (inertial scale), for $n = 4$. We observe that the structures are sheet/ribbon-like (b), or filament-like (c).}
	\label{3D}
\end{figure}

One of the main assets of DNS, is that they allow to study the three-dimensional organization of the local structures of cross-scale transfers. Fig. \ref{3D} displays the isosurfaces corresponding to $n = 4$ at $\ell/L = 0.125$ (inertial scale where nonlinear effects are the strongest), in a cubic volume corresponding to a quarter of the whole cube. For this set of parameters, the structures occupy close to $1\%$ of space, and contribute to approximately $10\%$ of the overall scale-to-scale energy transfer rate. We observe on Fig. \ref{3Dtout} that there exist structures of various sizes, almost all of them being elongated in one or two directions of space. Examples are displayed on Fig. \ref{nappe} and \ref{fils}. Fig. \ref{nappe} is obtained by zooming on the isosurface in the centre of Fig. \ref{3Dtout}. In this particular example, the length and width of the structure are of the same order of magnitude (they differ by a factor 2), but its thickness is around 60 times smaller than its length. This allows to qualify it as quasi 2D, or sheet-like. On the other hand, the structures displayed on Fig. \ref{fils} have a length approximately 10 times larger than their width, and can be considered as quasi 1D, or filament-like.

\begin{figure}	
	\centering
	\begin{subfigure}[t]{8.5cm}
		\caption{}
		\label{omegay}
		\begin{overpic}[width=8.5cm]{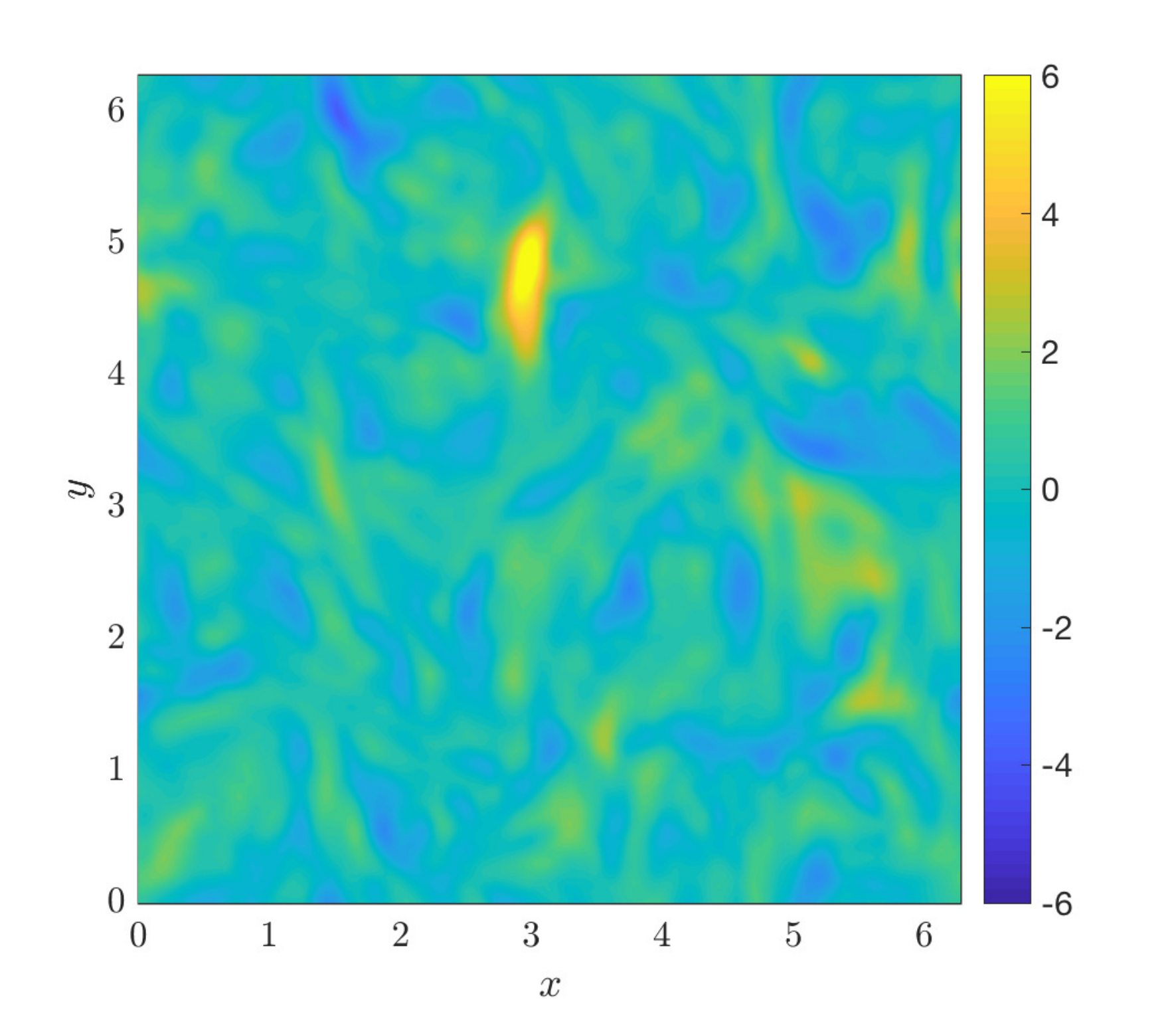}
       \put(70,83){\large $\omega_{y,\ell}$}
       \end{overpic}
	\end{subfigure}
	\quad
	\begin{subfigure}[t]{8.5cm}
		\caption{}
		\label{jz}
		\begin{overpic}[width=8.5cm]{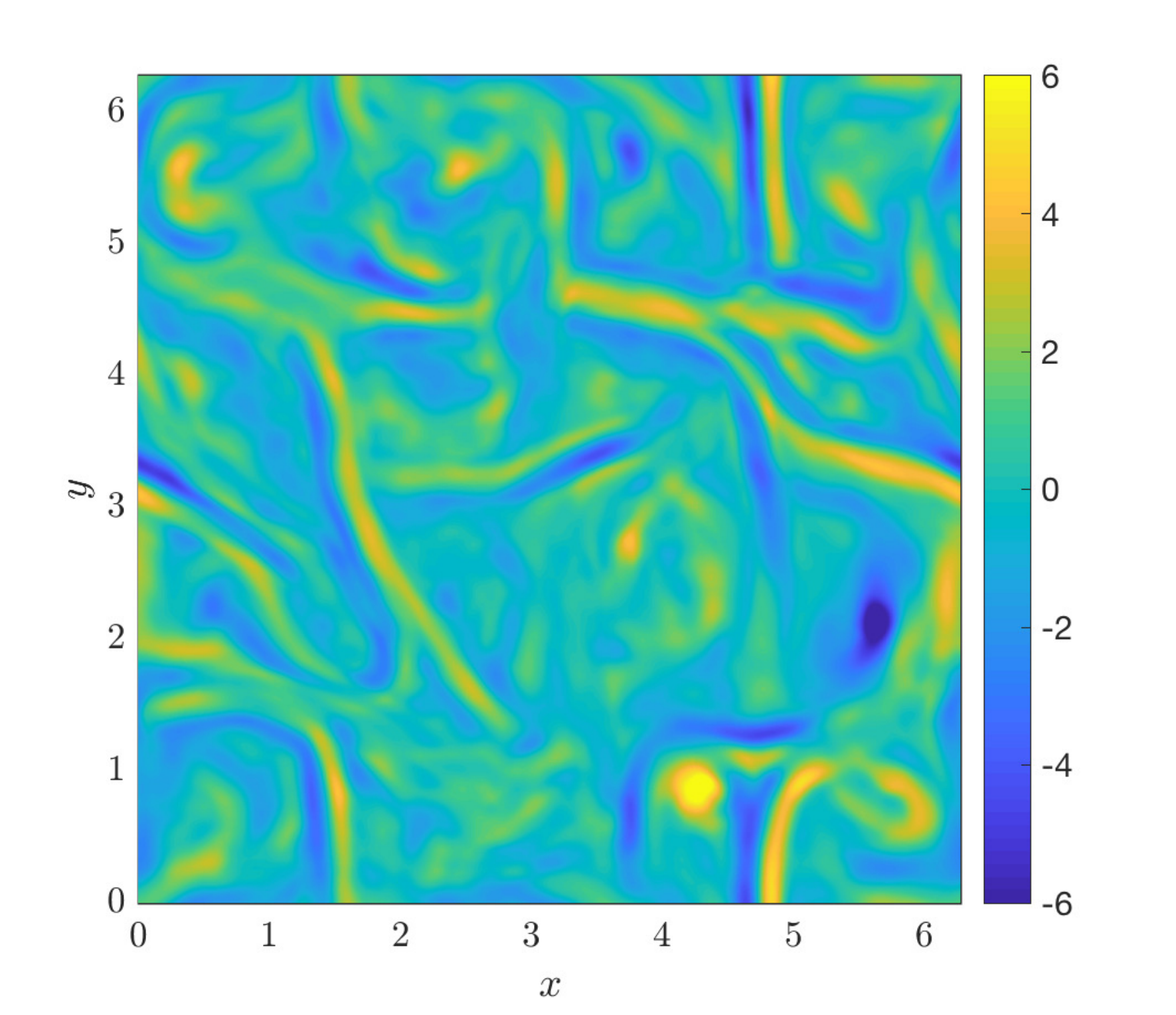}
       \put(20,83){\large $j_{z,\ell}$}
       \end{overpic}
	\end{subfigure}
	\caption{Slices of (a) the vorticity component $\omega_{y,\ell}$ and (b) the electric current density $j_{z,\ell}$ in the same plane as Fig. \ref{Fields} and \ref{TransfLoc}, at $\ell/L = 0.125$ (inertial scale). Comparing with Fig. \ref{TransfLocb}, we observe that structures of scale-to-scale transfers are located in regions of either high vorticity or vertical currents.}
	\label{Gradients}
\end{figure}

It seems natural to seek the origin of this three-dimensional organization in the topology of the velocity and magnetic fields. Indeed, the local transfers are determined by these two quantities, and a key question is to understand how the local structures of cross-scale transfers, observed for instance on Fig. \ref{TransfLoc}, are related to the local configuration of both $\bm u$ and $\bm b$ (displayed in the same plane on Fig. \ref{Fields}). It is known from hydrodynamics ($\bm b = 0$) that there exists a link between the finiteness of $\Pi^T_\ell$ at small scales and the possible existence of singularities in Navier-Stokes equations. In particular, for infinitely small scales, $\Pi^T_\ell$ converges (in the sense of distributions) to a scalar field which can only be nonzero in areas where $\bm u$ has infinite space gradients \cite{DR2000, Eyink2008}. The detection or study of such singularities in the MHD case is not the subject of this paper. However, as we noted in Sec. \ref{localOrg}, structures of cross-scale transfers exist on a wide range of scales and get more localized as $\ell$ is decreased. As a consequence, they might be the signature at finite scales of singularities in the velocity and magnetic fields. This idea has been investigated in the hydrodynamic case with encouraging results \cite{Saw2016, Kuzzay2017, MyThesis}. If this is actually correct, structures should appear in the same areas as strong velocity or magnetic field gradients, even after filtering. Fig. \ref{Gradients} displays two maps, in the same plane as in Fig. \ref{Fields} and \ref{TransfLoc}, of $\omega_{y,\ell} = \partial_z u_{x,\ell} - \partial_x u_{z,\ell}$ and $j_{z,\ell} = \partial_x b_{y,\ell} - \partial_y b_{x,\ell}$, at $\ell/L = 0.125$. Comparing with Fig. \ref{TransfLocb}, we see that areas of strong scale-to-scale transfers correlate well with areas of either strong vorticity or strong vertical current. Let us note that other components such as $j_{x,\ell}$, $j_{y,\ell}$, or $\omega_{x,\ell}$, do not display much correlation with local nonlinear transfers, or are highly correlated with the forcing, like $\omega_{z,\ell}$. The observed regions of strong vertical current are elongated over many $\left(xy\right)$ slices (several tens of them), so that what we see on Fig. \ref{jz} are vertical current sheets/ribbons or filaments. Therefore, vorticity and electric current could be used to localize areas of large cross-scale transfers. One might be surprised that we display curls to highlight areas of strong gradients instead of the gradients themselves. Indeed, each component of the curl is expressed as a difference of gradients, so that fields may have strong gradients and small curl. We have made this choice for mainly three reasons. First, in our study, we have not seen any regions of strong gradients with small curl. Second, vorticity and electric current density correspond to widely studied physical quantities which can be estimated from numerical as well as experimental data. Finally, recent studies have shown that when deformations of the pressure tensor are retained, they show high correlation with the vorticity \cite{Franci2016, del sarto2016}, suggesting that $\bm \omega$ can play a role in energy transfers when kinetic effect are implemented into the plasma description \cite{Parashar2016}. However, note that even if both $\bm \omega_\ell$ and $\bm j_\ell$ seem complementary when comparing with cross-scale transfers, it is interesting to note that in the case shown here, the electric current is much more correlated to $\Pi^T_\ell$ than the vorticity. This asymmetry may be due to the flow being forced only through Navier-Stokes equations, so that contrary to the electric current density, vorticity displays features of the forcing which can be observed on maps of $\omega_{z,\ell}$ even at inertial scales (figure not shown here). Whether this is a particular case or has more general implications needs further investigations.

\section{Conclusions}

In this article, we have shown the importance of investigating the local processes of nonlinear energy transfers in turbulent flows. Using a filtering approach, we are able to obtain a local Kármán-Howarth-Monin equation (\ref{localKHM}) which provides the exact expression for local cross-scale transfers at all scale $\ell$ (\ref{PiDR}), without assuming homogeneity nor isotropy. From this expression, a local form of Politano-Pouquet's 4/3-law can be derived (\ref{4/3local}), and the link with usual statistical results of turbulence can be established. The study of scale-to-scale energy transfers reveals that they fluctuate in space and organize into structures which can be locally extremely strong (more than one order of magnitude larger than the global cascade rate), even close to Kolmogorov scale. Some of these structures are observed over the whole available range of scales, and the computation of their filling factor shows that they get more localized as $\ell$ is decreased, while contributing to an increasing fraction of the cascade rate. In particular, we note that even close to Kolmogorov scale locally strong nonlinear effects persist. Finally, the three-dimensional study of these structures reveal that they are filament-like or sheet/ribbon-like, in agreement with their expected occurence in areas of strong gradients of $\bm u$ and $\bm b$. This last point is confirmed by comparing maps of local cross-scale transfers with maps of vorticity and electric current density.

The results presented in this paper open new perspectives for the study of the mechanisms leading up to the turbulent cascade. The local features of scale-to-scale transfers that we are able to highlight constitute a first step for future studies, which should allow us to gain new insights into the physics of turbulent MHD flows. As possible future investigations we can mention:

\begin{itemize}

\item MHD is characterized by a double direct cascade of both energy and cross helicity. In this paper, we have focused on the former in order to highlight the local (in space and time) features of the turbulent energy cascade. However, the cascade of cross helicity is another important mechanism of turbulent MHD flows, and should not be left aside. Performing the same study considering the local transfer rate $\Pi^C_\ell \coloneqq \left(\Pi^+_\ell - \Pi^-_\ell\right)/2$ of this quantity should yield a more complete understanding of the local processes generating the double MHD cascade.

\item Investigating other DNS with different characteristics. For instance, what happens in the presence of a mean magnetic field \cite{Verdini2015} or for two-dimensional turbulence ? 

\item A detailed analysis of the topologies of the velocity and magnetic fields at the location of strong cross-scale transfers should be performed in order to understand how the structures we observe emerge from the local configuration of the physical fields. Such a study may reveal local patterns producing these transfers such as what was observed in \cite{Saw2016, MyThesis}, and provide an explanation for their sign.

\item Studying the dynamics of these structures may also shed some lights on the local mechanisms generating scale-to-scale transfers. However, for three-dimensional turbulence with sufficiently high Reynolds numbers, this would require to process several time steps, which represents a huge amount of data and computational resources.

\item The existence of strong nonlinear effects close to Kolmogorov scale is an important result. Such locally extreme events have recently been studied in hydrodynamic turbulence \cite{Saw2016, MyThesis}, and their existence in MHD flows should be investigated. Indeed, a key question in plasma physics is whether the end of the MHD range is also the end of the cascade \cite{Leamon1998, Leamon1999, Leamon2000, Bale2005, Smith2006}, or whether there is some energy which continues to be transferred to kinetic scales as suggested in \cite{Yang2017, Biskamp1996, Ghosh1996, Li2001, Stawicki2001, Galtier2006, Alexandrova2008}. Recent studies have shown a possible link between various energy transfer channels, based on proxies for the local energy cascade rate and for energy dissipation \cite{Yang2018}. Applying the more rigorous approach presented in this paper to kinetic simulations and Hall-MHD should help understanding what happens for energy transfers at scales where kinetic physics becomes important, especially since the local Hall-MHD term has been derived in \cite{Galtier2018}.

\item Another subject worth exploring is the relation between the nonlinear and viscous terms. In particular, small-scale dynamics is mostly governed by viscous forces, and it seems natural to investigate whether structures of cross scale transfers exist in areas of enhanced viscous dissipation, or are completely uncorrelated. In the MHD regime, it might be expected that since locally strong scale-to-scale transfers appear in areas of strong gradients, these gradients also generate strong viscous dissipation, therefore inducing a good correlation. However, this is without taking the smallness of the viscosity into account. Note that previous studies at kinetic scales using partial variance of increments (PVI) show that temperature anisotropy and particle distribution function distortions, probably due to energy dissipation, are concentrated near structures such as current sheets \cite{Greco2012}. The relation between nonlinear and viscous terms has been investigated at various scales in hydrodynamic turbulence, and results suggest that the extreme events of interscale transfer correspond to the minima of viscous dissipation, and vice versa \cite{Debue2018}. The same question for plasma turbulence, however, is still to be addressed.

\item As we mentioned in Sec. \ref{localbalance}, it is possible to derive two local balances for the kinetic and magnetic energies separately (see App. \ref{pourAluie}). To our knowledge, this is the first time that such balances are derived using the point-split approach. However, their counterparts using the LES point of view are known and have been investigated \cite{Aluie2010, Aluie2017, Yang2017,Camporeale2018, Bian2019}. As a consequence, these results will allow to compare the two approaches, and understand the local processes governing the transfers of kinetic and magnetic energies separately. For instance, recent studies using DNS of MHD turbulence indicate that kinetic/magnetic energy exchanges only occur at large scales on average \cite{Bian2019}. This means that the two energy budgets statistically decouple at intermediate and small scales, and suggests that the cascade phenomenology applies to kinetic and magnetic energies separately, even though they are not quadratic invariants. Formalizing such a phenomenology in the spirit of K41 may therefore be possible. Additionally, the authors find that the two cascade rates are equal. Checking whether these results hold in the point-split approach, based on App. \ref{pourAluie}, could help supplement this work.

\item The study presented in this paper is also particularly relevant for solar wind physics where turbulence seems to be a key mechanism to explain solar wind heating, observed from the slow decrease of its temperature with distance from the Sun. Applying the approach presented in this paper to spacecraft data should therefore allow to understand the mechanisms at work in the solar wind.

\item Finally, let us mention the case of compressible turbulence, where scale-filtering has been used to investigate the scale locality of the cascade using the LES approach, and for which the point-split technique should be worth investigating \cite{Aluie2011, Yang2016}.

\end{itemize}

\section{Acknowledgements}

The work of D. Kuzzay is supported by the LABEX PLAS@PAR and the French Centre National d'Etude Spatiales (CNES). L. Matteini acknowledges the support of the Programme National PNST of CNRS/INSU co-funded by CNES. Olga Alexandrova is supported CNES.

\newpage

\appendix

\section{Derivation of separate balances for large-scale kinetic and magnetic energies}
\label{pourAluie}

In this appendix, we derive two balances for large-scale kinetic and magnetic energies separately, using the point-split approach. They are given in Eq. (\ref{KEbalanceCG}) and (\ref{MEbalanceCG}), and in an alternative form in Eq. (\ref{KEbalanceCG2}) and (\ref{MEbalanceCG2}). The derivation is largely inspired by \cite{EyinkNotes}. We start from the incompressible MHD equations at point $\bm x$

\begin{align}
\label{MHDeqUx}
\partial_t u_i + u_j \partial_j u_i &= -\partial_i p^\ast + b_j \partial_j b_i + \nu \partial_j\partial_j u_i,\\
\label{incompUx}
\partial_j u_j &= 0,\\
\label{MHDeqBx}
\partial_t b_i + u_j \partial_j b_i &= b_j \partial_j u_i + \eta \partial_j\partial_j b_i,\\
\label{incompBx}
\partial_j b_j &= 0,
\end{align}

and point $\bm x + \bm r$

\begin{align}
\label{MHDeqUxr}
\partial_t \widehat{u_i} + \widehat{u_j} \partial_j \widehat{u_i} &= -\partial_i \widehat{p^\ast} + \widehat{b_j} \partial_j \widehat{b_i} + \nu \partial_j\partial_j \widehat{u_i},\\
\label{incompUxr}
\partial_j \widehat{u_j} &= 0,\\
\label{MHDeqBxr}
\partial_t \widehat{b_i} + \widehat{u_j} \partial_j \widehat{b_i} &= \widehat{b_j} \partial_j \widehat{u_i} + \eta \partial_j\partial_j \widehat{b_i},\\
\label{incompBxr}
\partial_j \widehat{b_j} &= 0.
\end{align}

The idea is the same as what we have done in the paper. We will derive a balance for the point-split energy, and then apply a filter in order to obtain a coarse-grained balance at scale $\ell$. We will start with the kinetic energy (KE). Let us take the dot product of Eq. (\ref{MHDeqUx}) with $\widehat{\bm u}$ together with the dot product of Eq. (\ref{MHDeqUxr}) with $\bm u$. We get

\begin{align}
\widehat{u_i}\partial_t u_i + \widehat{u_i}\partial_j \left(u_i u_j\right) &= -\partial_i\left(\widehat{u_i} p^\ast\right) + \widehat{u_i} \partial_j \left(b_i b_j\right) + \nu \widehat{u_i}\partial_j\partial_j u_i,\\
u_i\partial_t \widehat{u_i} + u_i\partial_j \left(\widehat{u_i}\widehat{u_j}\right) &= -\partial_i\left(u_i \widehat{p^\ast}\right) + u_i\partial_j \left(\widehat{b_i}\widehat{b_j} \right) + \nu u_i\partial_j\partial_j \widehat{u_i},
\end{align}

Summing these two equations we obtain

\begin{equation}
\label{toto}
\underbrace{\partial_t\left(u_i \widehat{u_i}\right)}_{\text{time derivative}} + \underbrace{u_i\partial_j\left(\widehat{u_i}\widehat{u_j} - \widehat{b_i}\widehat{b_j}\right) + \widehat{u_i} \partial_j \left(u_i u_j - b_i b_j\right)}_{\text{nonlinear terms}} = \underbrace{-\partial_i\left( u_i \widehat{p^\ast} + \widehat{u_i} p^\ast \right)}_{\text{pressure terms}} + \underbrace{\nu\left( u_i \partial_j\partial_j \widehat{u_i} + \widehat{u_i} \partial_j\partial_j u_i\right)}_{\text{viscous terms}}.
\end{equation}

We have put time derivatives and nonlinear terms together on the left hand side, and pressure as well as viscous terms together on the right hand side. Now we are going to treat the viscous and nonlinear terms separately.\\

$\bullet$ The viscous term can be rewritten

\begin{equation*}
u_i \partial_j\partial_j \widehat{u_i} + \widehat{u_i} \partial_j\partial_j u_i = \partial_j \partial_j \left(u_i \widehat{u_i}\right) - 2\left(\partial_ju_i\right)\left(\partial_j \widehat{u_i}\right)
\end{equation*}

$\bullet$ We know from \cite{EyinkNotes} (chapter III C) that nonlinear terms containing only the velocity field can be written as a term describing scale-to-scale transfers of KE plus a divergence. Those mixing both the velocity and magnetic fields are left untouched for now and will be treated later.

\begin{equation*}
u_i\partial_j\left(\widehat{u_i}\widehat{u_j}\right) + \widehat{u_i} \partial_j \left(u_i u_j \right) = -\underbrace{\frac{1}{2}\partial_{r_j}\left( \delta u_j \delta u_i \delta u_i\right)}_{\Pi^u} + \partial_j \left( \frac{1}{2}\widehat{u_i}\widehat{u_i} \delta u_j + u_i \widehat{u_i} u_j \right)
\end{equation*}

We can then write Eq. (\ref{toto}) as

\begin{multline}
\label{titi}
\partial_t\left(u_i \widehat{u_i}\right) + \partial_j \left( u_i \widehat{u_i} u_j + \frac{1}{2}\widehat{u_i}\widehat{u_i} \delta u_j +  u_j \widehat{p^\ast} + \widehat{u_j} p^\ast - \nu \partial_j\left( u_i \widehat{u_i} \right)\right)\\
= \underbrace{u_i\partial_j \left(\widehat{b_i}\widehat{b_j} \right)}_{\textcircled{1}} + \underbrace{\widehat{u_i} \partial_j \left(b_i b_j\right)}_{\textcircled{2}} + \Pi^u - 2 \nu\left(\partial_ju_i\right)\left(\partial_j \widehat{u_i}\right),
\end{multline}

where \textcircled{1} and \textcircled{2} are the terms which have not yet been treated.\\

We now move on to the point-split magnetic energy (ME) balance. The calculations are the same as what we have done for the KE. Following the same procedure with Eq. (\ref{MHDeqBx}) and (\ref{MHDeqBxr}) we get

\begin{align}
\widehat{b_i}\partial_t b_i + \widehat{b_i}\partial_j \left(b_i u_j\right) &= \widehat{b_i}\partial_j \left(u_i b_j\right)  + \eta \widehat{b_i}\partial_j\partial_j b_i,\\
b_i\partial_t \widehat{b_i} + b_i\partial_j \left(\widehat{b_i}\widehat{u_j}\right) &= b_i\partial_j \left(\widehat{u_i}\widehat{b_j}\right) + \eta b_i\partial_j\partial_j \widehat{b_i},
\end{align}

We sum these two equations 

\begin{equation}
\underbrace{\partial_t\left(b_i \widehat{b_i}\right)}_{\text{time derivative}} + \underbrace{b_i\partial_j\left(\widehat{b_i}\widehat{u_j} - \widehat{u_i}\widehat{b_j}\right) + \widehat{b_i} \partial_j \left(b_i u_j - u_i b_j\right)}_{\text{nonlinear terms}} = \underbrace{\eta\left( b_i \partial_j\partial_j \widehat{b_i} + \widehat{b_i} \partial_j\partial_j b_i\right)}_{\text{dissipative terms}},
\end{equation}

and after performing the same rearrangements as for the KE balance we get

\begin{equation}
\label{tata}
\partial_t\left(b_i \widehat{b_i}\right) + \partial_j \left( b_i \widehat{b_i} u_j + \frac{1}{2}\widehat{b_i}\widehat{b_i} \delta u_j - \eta \partial_j\left( b_i \widehat{b_i} \right)\right) = \underbrace{b_i\partial_j \left(\widehat{u_i}\widehat{b_j} \right)}_{\textcircled{3}} + \underbrace{\widehat{b_i} \partial_j \left(u_i b_j\right)}_{\textcircled{4}} + \Pi^b - 2 \eta\left(\partial_jb_i\right)\left(\partial_j \widehat{b_i}\right).
\end{equation}

Again, \textcircled{3} and \textcircled{4} are the terms which have yet to be treated, and we defined $\Pi^b \coloneqq \frac{1}{2}\partial_{r_j}\left( \delta u_j \delta b_i \delta b_i\right)$.\\

Let us now search for a relation which links the four terms \textcircled{1}, \textcircled{2}, \textcircled{3} and \textcircled{4}. First, we note that

\begin{align*}
\textcircled{1} + \textcircled{4} &= \underbrace{u_i \widehat{b_j} \partial_j\widehat{b_i} - u_i b_j\partial_j\widehat{b_i}}_{\Delta_1} + \partial_j\left( u_i \widehat{b_i}b_j\right) = \underbrace{u_i\delta b_j \partial_j \widehat{b_i}}_{\Delta_1} + \partial_j\left( u_i \widehat{b_i}b_j \right),\\
\textcircled{2} + \textcircled{3} &= \underbrace{b_i \widehat{b_j} \partial_j\widehat{u_i} - b_i b_j\partial_j\widehat{u_i}}_{\Delta_2} + \partial_j\left( \widehat{u_i} b_i b_j\right) = \underbrace{b_i\delta b_j \partial_j \widehat{u_i}}_{\Delta_2} + \partial_j\left( \widehat{u_i} b_i b_j\right).
\end{align*}

Here again we have followed the steps of \cite{EyinkNotes} (chapter III C). Second,

\begin{align*}
\partial_{r_j} \left(\delta b_j \delta u_i \delta b_i\right) &= \delta b_j\partial_{r_j}\left(\delta u_i \delta b_i\right)\\
&= \delta b_j\left( \delta b_i \partial_{r_j} \delta u_i + \delta u_i \partial_{r_j} \delta b_i \right)\\
& = \delta b_j\left( \delta b_i \partial_j \widehat{u_i} + \delta u_i \partial_j \widehat{b_i} \right)\\
&= \delta b_j \left( \widehat{b_i} \partial_j \widehat{u_i} - b_i\partial_j \widehat{u_i} + \widehat{u_i} \partial_j \widehat{b_i} - u_i\partial_j \widehat{b_i} \right)\\
&= \delta b_j \partial_j\left( \widehat{u_i} \widehat{b_i}\right) - \Delta_1 - \Delta_2\\
&= \partial_j\left( \delta b_j \widehat{u_i} \widehat{b_i}\right) - \Delta_1 - \Delta_2
\end{align*}

We therefore deduce the relation

\begin{equation}
\textcircled{1} + \textcircled{2} + \textcircled{3} + \textcircled{4} = \underbrace{- \partial_{r_j} \left( \delta b_j \delta u_i \delta b_i \right)}_{\Pi^{u,b}} + \partial_j \left( \delta b_j \widehat{u_i} \widehat{b_i} + u_i \widehat{b_i}b_j + \widehat{u_i} b_i b_j \right)
\label{ImportantRelation}
\end{equation}

Assembling Eq. (\ref{titi}), (\ref{tata}), and (\ref{ImportantRelation}), we can write

\begin{align}
\begin{split}
\partial_t\left(u_i \widehat{u_i}\right) + &\partial_j \left( u_i \widehat{u_i} u_j + \frac{1}{2}\widehat{u_i}\widehat{u_i} \delta u_j +  u_j \widehat{p^\ast} + \widehat{u_j} p^\ast - \nu \partial_j\left( u_i \widehat{u_i} \right) - u_i \widehat{b_i} \widehat{b_j} -\widehat{u_i} b_i b_j \right)\\
 &= \Pi^u - \widehat{b_i}\widehat{b_j} \partial_j u_i - b_i b_j \partial_j \widehat{u_i} - 2 \nu\left(\partial_ju_i\right)\left(\partial_j \widehat{u_i}\right),
 \end{split}
 \\
 \begin{split}
\partial_t\left(b_i \widehat{b_i}\right) + &\partial_j \left( b_i \widehat{b_i} u_j + \frac{1}{2}\widehat{b_i}\widehat{b_i} \delta u_j - \eta \partial_j\left( b_i \widehat{b_i} \right) - \delta b_j \delta u_i \widehat{b_i} \right)\\
&= \Pi^b + \Pi^{u,b} + \widehat{b_i}\widehat{b_j} \partial_j u_i + b_i b_j \partial_j \widehat{u_i} - 2 \eta\left(\partial_jb_i\right)\left(\partial_j \widehat{b_i}\right).
\end{split}
\end{align}

Multiplying these two identities by the kernel $G_\ell\left(\bm r\right)$ and integrating over $\bm r$ yields the coarse-grained KE and ME balances

\begin{align}
\begin{split}
\partial_t\left(\frac{1}{2}u_i u_{i,\ell}\right) + &\partial_j \left( \frac{1}{2}u_i u_{i,\ell} u_j + \frac{1}{4}\left(u_i u_i u_j \right)_\ell - \frac{1}{4} \left(u_i u_i \right)_\ell u_j +  \frac{1}{2} \left(u_j p^\ast_\ell + u_{j,\ell} p^\ast\right) - \nu \partial_j\left( \frac{1}{2}u_i u_{i,\ell} \right) -\frac{1}{2} \left( u_i \left(b_i b_j\right)_\ell + u_{i,\ell} b_i b_j\right) \right)\\
 &= -\Pi^u_\ell - \frac{1}{2} \left(\left(b_i b_j\right)_\ell \partial_j u_i + b_i b_j \partial_j u_{i,\ell}\right) - \nu\left(\partial_ju_i\right)\left(\partial_j u_{i,\ell}\right),
 \end{split}
 \label{KEbalanceCG}
 \\
 \begin{split}
\partial_t\left(\frac{1}{2}b_i b_{i,\ell}\right) + &\partial_j \left( \frac{1}{2}b_i b_{i,\ell} u_j + \frac{1}{4}\left(b_i b_i u_j \right)_\ell - \frac{1}{4} \left(b_i b_i \right)_\ell u_j - \eta \partial_j\left( \frac{1}{2}b_i b_{i,\ell} \right) - \frac{1}{2} \left( \left(u_i b_i b_j\right)_\ell -\left( b_i b_j\right)_\ell u_i - \left(u_i b_i\right)_\ell b_j + u_i b_{i,\ell} b_j \right)\right)\\
&= -\Pi^b_\ell - \Pi^{u,b}_\ell + \frac{1}{2} \left(\left(b_i b_j\right)_\ell \partial_j u_i + b_i b_j \partial_j u_{i,\ell}\right) - \eta\left(\partial_jb_i\right)\left(\partial_j b_{i,\ell}\right),
\end{split}
\label{MEbalanceCG}
\end{align}

where

\begin{align*}
\Pi^u_\ell &\coloneqq \frac{1}{4} \int \ d\bm r \left[\partial_j G_\ell \left(\bm r\right)\right] \delta u_j \delta u_i \delta u_i,\\
\Pi^b_\ell &\coloneqq \frac{1}{4} \int \ d\bm r \left[\partial_j G_\ell \left(\bm r\right)\right] \delta u_j \delta b_i \delta b_i,\\
\Pi^{u,b}_\ell &\coloneqq - \frac{1}{4} \int \ d\bm r \left[\partial_j G_\ell \left(\bm r\right)\right] 2\delta b_j \delta u_i \delta b_i,
\end{align*}

Of course, summing Eq. (\ref{KEbalanceCG}) and (\ref{MEbalanceCG}) yields the local KHM relation (\ref{localKHM}). Let us note that $\Pi^u_\ell$, $\Pi^b_\ell$ and $\Pi^{u,b}_\ell$ are the local form of the three terms which appear in Politano and Pouquet's 4/3-law \cite{PP1998b}. We also note that an additional term describing the exchanges between KE and ME appears, and take the form $\left(\left(b_i b_j\right)_\ell \partial_j u_i + b_i b_j \partial_j u_{i,\ell}\right)/2$. Since it appears with opposite signs in the two balances, it does not contribute to the local balance for the total energy. Using the strain tensor $S_{ij} = \left( \partial_i u_j + \partial_j u_i \right)/2$, this term can be rewritten as $\left(\left(b_i b_j\right)_\ell S_{ij} + b_i b_j S^\ell_{ij} \right)/2$. In this form, it is interpreted as the kinetic energy expended by the flow to bend and stretch the magnetic field lines (see \cite{Aluie2017} for the same discussion in the large-eddy simulation approach). Alternatively, it can be related to the work of the Lorentz force $\bm f_L$ on the particles. Denoting by $\bm j$ the electric current density we get

\begin{align*}
\left(\widehat{\bm j} \times \widehat{\bm b}\right) \cdot \bm u &= \left[\left(\bm\nabla \times \widehat{\bm b}\right) \times \widehat{\bm b}\right]\cdot \bm u,\\
&= \left[\left(\widehat{\bm b} \cdot \bm \nabla\right) \widehat{\bm b} - \bm \nabla\left( \frac{1}{2}\vert \widehat{\bm b}\vert^2 \right)\right] \cdot \bm u,\\
& = u_i \partial_j \left( \widehat{b_i}\widehat{b_j}\right) -\partial_j\left( \frac{1}{2} \widehat{b_i}\widehat{b_i} u_j \right),\\
&= -\widehat{b_i}\widehat{b_j}\partial_j u_i + \partial_j\left( u_i \widehat{b_i}\widehat{b_j} -   \frac{1}{2} \widehat{b_i}\widehat{b_i} u_j \right),
\end{align*}

which after filtering gives

\begin{equation*}
f^\ell_{L,i} u_i = -\left(b_i b_j\right)_\ell \partial_j u_i + \partial_j \left( u_i \left(b_i b_j\right)_\ell - \frac{1}{2} \left(b_i b_i\right)_\ell u_j \right).
\end{equation*}

In the same way

\begin{equation*}
f_{L,i} u_{i,\ell} = - b_i b_j \partial_j u_{i,\ell} + \partial_j \left( u_{i,\ell} b_i b_j - \frac{1}{2} b_i b_i u_{j,\ell} \right),
\end{equation*}

and we recover the interpretation of KE/ME exchanges due to the interactions of the plasma with the electromagnetic fields through the Lorentz force (see also \cite{Aluie2017}). Eq. (\ref{KEbalanceCG}) and (\ref{MEbalanceCG}) can then be rewritten

\begin{align}
\begin{split}
\partial_t\left(\frac{1}{2}u_i u_{i,\ell}\right) + &\partial_j \left( \frac{1}{2}u_i u_{i,\ell} u_j + \frac{1}{4}\left(u_i u_i u_j \right)_\ell - \frac{1}{4} \left(u_i u_i \right)_\ell u_j +  \frac{1}{2} \left(u_j p^\ast_\ell + u_{j,\ell} p^\ast\right) - \nu \partial_j\left( \frac{1}{2}u_i u_{i,\ell} \right) - \frac{1}{4} \left( \left(b_i b_i\right)_\ell u_j + b_i b_i u_{j,\ell} \right) \right)\\
 &= -\Pi^u_\ell + \frac{1}{2} \left(f^\ell_{L,i} u_i + f_{L,i} u_{i,\ell} \right) - \nu\left(\partial_ju_i\right)\left(\partial_j u_{i,\ell}\right),
 \end{split}
  \label{KEbalanceCG2}
 \\
 \begin{split}
\partial_t\left(\frac{1}{2}b_i b_{i,\ell}\right) + &\partial_j \left( \frac{1}{2}b_i b_{i,\ell} u_j + \frac{1}{4}\left(b_i b_i u_j\right)_\ell - \eta \partial_j\left( \frac{1}{2}b_i b_{i,\ell} \right) - \frac{1}{2} \left( \left(u_i b_i b_j\right)_\ell - \left(u_i b_i\right)_\ell b_j + u_i b_{i,\ell} b_j \right) - \frac{1}{2} u_{i,\ell} b_i b_j +  \frac{1}{4} b_i b_i u_{j,\ell} \right)\\
&= -\Pi^b_\ell - \Pi^{u,b}_\ell - \frac{1}{2} \left(f^\ell_{L,i} u_i + f_{L,i} u_{i,\ell} \right) - \eta\left(\partial_jb_i\right)\left(\partial_j b_{i,\ell}\right).
\end{split}
 \label{MEbalanceCG2}
\end{align}

\end{document}